\documentclass[journal,twocolumn,10pt,twoside]{IEEEtran}
\usepackage{amsmath,amsopn,amssymb,bm}
\usepackage{gensymb}
\usepackage{amsfonts}
\usepackage{amsthm} 
\usepackage{graphicx,xspace,color}
\usepackage{epsfig}
\usepackage{longtable,multirow}
\usepackage{array}
\usepackage{stfloats}
\usepackage{cuted}
\usepackage{cite}
\usepackage[nolist]{acronym}
\usepackage{soul}
\usepackage{algorithm}
\usepackage{algpseudocode}
\usepackage{enumerate}
\usepackage{lettrine}
\usepackage{mathtools}
\usepackage{comment}
\usepackage{subcaption}

\usepackage[font=footnotesize]{caption}

\usepackage{booktabs}
\graphicspath{ {./figures/}}

\newtheorem{remark}{Remark}

\newtheorem{proposition}{Proposition}
\newtheorem{lemma}{Lemma}

\begin{document}
	
\begin{acronym}
	\acro{THz}{Terahertz}
	\acro{mmWave}{millimeter wave}
	\acro{OFDM}{orthogonal frequency division multiplexing}
	\acro{SFW}{spatial-frequency wideband}
	\acro{UPA}{uniform planar array}
	\acro{ULAs}{uniform linear arrays}
	\acro{BS}{base station}
	\acro{TTD}{true-time-delay}
	\acro{CS}{compressive sensing}
	\acro{OMP}{orthogonal matching pursuit}
	\acro{GSOMP}{generalized simultaneous OMP}
	\acro{SNR}{signal-to-noise ratio}
	\acro{CRLB}{Cram\'{e}r-Rao lower bound}
	\acro{RF}{radio frequency}
	\acro{DoA}{direction-of-arrival}
	\acro{DoD}{direction-of-departure}
	\acro{ToA}{time-of-arrival}
	\acro{MIMO}{multiple-input multiple-output}
	\acro{LS}{least squares}
	\acro{CSI}{channel state information}
	\acro{LoS}{line-of-sight}
	\acro{NLoS}{non-line-of-sight}
	\acro{DFT}{discrete Fourier transform}
	\acro{CDF}{cumulative distribution function}
	\acro{AoD}{angle-of-departure}
	\acro{PAPR}{peak-to-average power ratio}
	\acro{NMSE}{normalized mean-square error}
	\acro{SVD}{singular value decomposition}
	\acro{MRC}{maximum-ratio combiner}
	\acro{3D}{three-dimensional}
\end{acronym}

\title{Channel Estimation and Hybrid Combining for Wideband Terahertz Massive MIMO Systems}

\author{Konstantinos Dovelos, Michail Matthaiou, Hien Quoc Ngo, and Boris Bellalta

\thanks{K. Dovelos and B. Bellalta are with the Department
	of Information and Communication Technologies, Pompeu Fabra University, Barcelona, Spain (e-mail: konstantinos.dovelos@upf.edu; boris.bellalta@upf.edu).}

\thanks{M. Matthaiou and H. Q. Ngo are with the Institute of Electronics, Communications and Information Technology (ECIT), Queen’s University Belfast, Belfast, U.K. (e-mail: m.matthaiou@qub.ac.uk; hien.ngo@qub.ac.uk).}
}

\maketitle
\begin{abstract}
\ac{THz} communication is widely considered as a key enabler for future 6G wireless systems. However, \ac{THz} links are subject to high propagation losses and inter-symbol interference due to the frequency selectivity of the channel. Massive \ac{MIMO} along with \ac{OFDM} can be used to deal with these problems. Nevertheless, when the propagation delay across the \ac{BS} antenna array exceeds the symbol period, the spatial response of the \ac{BS} array varies across the \ac{OFDM} subcarriers. This phenomenon, known as beam squint, renders narrowband combining approaches ineffective. Additionally, channel estimation becomes challenging in the absence of combining gain during the training stage. In this work, we address the channel estimation and hybrid combining problems in wideband \ac{THz} massive \ac{MIMO} with uniform planar arrays. Specifically, we first introduce a low-complexity beam squint mitigation scheme based on true-time-delay. Next, we propose a novel variant of the popular \ac{OMP} algorithm to accurately estimate the channel with low training overhead. Our channel estimation and hybrid combining schemes are analyzed both theoretically and numerically. Moreover, the proposed schemes are extended to the multi-antenna user case. Simulation results are provided showcasing the performance gains offered by our design compared to standard narrowband combining and OMP-based channel estimation.
\end{abstract}


\section{Introduction}
Spectrum scarcity is the main bottleneck of current wireless systems. As a result, new frequency bands and signal processing techniques are needed to deal with this spectrum gridlock. In view of the enormous bandwidth available at \acf{THz} frequencies, communication over the \ac{THz} band is considered as a key technology for future 6G wireless systems~\cite{6G_Survey}. More particularly, the \ac{THz} band, spanning from $0.1$ to $10$ THz, offers bandwidths orders of magnitude larger than the \ac{mmWave} band. For example, the licensed bandwidth in the \ac{mmWave} band is usually up to $7$~GHz whilst that of the THz band is at least $10$~GHz~\cite{towards_thz_com}. On the other hand, as the frequency increases, the signals experience much more severe path attenuation compared to their mmWave and microwave counterparts, according to Friis transmission formula. Thanks to the very short wavelength of THz signals, though, a very large number of antennas can be tightly packed into a small area to form a massive \acf{MIMO} transceiver, and effectively compensate for the propagation losses by means of beamforming~\cite{prospective_mimo}. Therefore, \ac{THz} massive MIMO is expected to be a key enabler for ultra-high-speed networks, such as terabit wireless personal/local area networks and femtocells~\cite{future_of_thzcom}.

Despite the promising performance gains of \ac{THz} massive MIMO systems, the wideband transmissions in conjunction with the large array aperture, with respect to the symbol period, give rise to \textit{\ac{SFW}} effects~\cite{spatial_freq_wideband_effects}. Specifically, the channel response becomes frequency-selective not only because of the delay spread of the multi-path channel, but also due to the large propagation delay across the array aperture~\cite{spatial_wideband_effect}. As a result, the response of the \ac{BS} array can be frequency-dependent also in a \ac{LoS} scenario. When \acf{OFDM} modulation is employed to combat inter-symbol interference, the spatial-wideband effect renders the \ac{DoA} and \ac{DoD} of the signals to vary across the subcarriers. This phenomenon, termed \textit{beam squint}, calls for frequency-dependent beamforming/combining, which is not available in a typical hybrid array architecture of \ac{THz} massive MIMO. More particularly, narrowband beamforming/combining approaches can substantially reduce the array gain across the \ac{OFDM} subcarriers, hence leading to performance degradation~\cite{phased_array_handbook}. Consequently, beam squint compensation is of paramount importance for \ac{THz} massive \ac{MIMO}-\ac{OFDM} systems.

Since accurate \ac{CSI} is essential to effectively apply combining and/or beam squint mitigation, channel estimation under \ac{SFW} effects is another important problem to address. Specifically, in the absence of combining gain during channel estimation, the detection of the paths present in the channel becomes challenging in the low \ac{SNR} regime. Additionally, due to the massive number of \ac{BS} antennas and the limited number of \ac{RF} chains in a hybrid array architecture, the channel estimation overhead becomes excessively large even for single-antenna users under standard approaches, such as the~\ac{LS} method. In conclusion, \ac{THz} massive MIMO brings new challenges in the signal processing design, and calls for carefully tailored solutions that take into account the unique propagation characteristics in \ac{THz} bands. 

\subsection{Prior Work}
In this section, we review prior work on channel estimation and hybrid beamforming in wideband \ac{mmWave}/\ac{THz} systems.

The authors in~\cite{single_carrier_THz} proposed a novel single-carrier transmission scheme for \ac{THz} massive MIMO, which utilizes minimum mean-square error precoding and detection. Nevertheless, a narrowband antenna aray model was considered, and hence the \ac{SFW} effect was ignored. A stream of recent papers on wideband \ac{mmWave} MIMO-OFDM systems (see \cite{mmWave_bsq_ref0, mmWave_bsq_ref1,mmWave_bsq_ref2, mmWave_bsq_ref3}, and references therein) proposed methods to jointly optimize the analog combiner and the digital precoder in order to maximize the achievable rate under the beam squint effect. In a similar spirit, \cite{mmWave_bsq_ref4} and \cite{mmWave_bsq_ref5} proposed a new analog beamforming codebook with wider beams to avoid the array gain degradation due to beam squint. These methods can enhance the achievable rate when the beam squint effect is mild. However, their performance becomes poor in \ac{THz} \ac{MIMO} systems due to the much larger signaling bandwidth and number of \ac{BS} antennas compared to their \ac{mmWave} counterparts~\cite{Hybrid_TD_precoding}. To this end, \cite{AosA_TD_mmWave} proposed a wideband codebook for beam training for \ac{ULAs} using~\ac{TTD}~\cite{true_tdl}. However, this design is limited to~\ac{ULAs} and beam alignment without explicitely estimating the channel. From the relevant literature on hybrid beamforming, we distinguish~\cite{Hybrid_TD_precoding}, which proposed a \ac{TTD}-based hybrid beamformer for \ac{THz} massive MIMO, however assuming \ac{ULAs} and perfect \ac{CSI}. 

Despite the importance of channel estimation, there are only few recent works in the literature investigating the channel estimation problem under the spatial-wideband effect. More particularly, the seminal paper~\cite{spatial_freq_wideband_effects} introduced the \ac{SFW} for \ac{mmWave} massive MIMO systems, and proposed a channel estimation algorithm by capitalizing on the asymptotic properties of \ac{SFW} channels. However, the proposed algorithm relies on multiplying the channel of an $N$-element uniform linear array by an $N$-point \ac{DFT} matrix, and hence entails high training overhead when the number of \ac{RF} chains is much smaller than the number of \ac{BS} antennas. In a similar spirit,~\cite{power_focusing_approach} employed the \acf{OMP} algorithm along with an energy-focusing preprocessing step to estimate the \ac{SFW} channel, while minimizing the power leakage effect. Finally,~\cite{ce_frequency_selective_mmWave} leveraged tools from \ac{CS} theory to tackle the channel estimation problem in frequency-selective multiuser \ac{mmWave} \ac{MIMO} systems but in the absence of the spatial-wideband effect. 
\subsection{Contributions}
 In this paper, we address the channel estimation and hybrid combining problems in wideband THz MIMO. To this end, we assume \ac{OFDM} modulation, which is the most popular transmission scheme over frequency-selective channels. The main contributions of the paper are summarized as follows:
\begin{itemize}
\item 
We model the \ac{SFW} effect in THz MIMO-\ac{OFDM} systems with a \ac{UPA} at the \ac{BS}. Note that prior studies (e.g.,~\cite{indoor_thz, THz_spatial_modulation}) on mmWave/THz communication with UPAs ignore the \ac{SFW} effect. We next show that frequency-flat combining leads to substantial performance losses due to the severe beam squint effect occuring across \ac{OFDM} subcarriers, and propose a beam squint compensation strategy using \ac{TTD}~\cite{ttd_power_cons_ref2} and virtual array partition. The scope of the virtual array partition is to reduce the number of \ac{TTD} elements needed to effectively mitigate beam squint. To this end, we derive the wideband combiner expression for a rectangular planar array, and establish its near-optimal performance with respect to fully-digital combining analytically, as well as through computer simulations. 

\item We propose a solution to the channel estimation problem under the \ac{SFW} effect. Specifically, by availing of the channel sparsity in the angular domain, we first adopt a sparse representation of the THz channel, and formulate the channel estimation problem as a \ac{CS} problem. We then propose a solution based on the \ac{OMP} algorithm, which is one of the most common and simple greedy \ac{CS} methods. Contrary to existing works, we employ a \textit{wideband} dictionary and show that channels across different \ac{OFDM} subcarriers share a common support. This enables us to apply a variant of the simultaneous \ac{OMP} algorithm, coined as \ac{GSOMP}, which exploits the information of multiple subcarriers to increase the probability of successfully recovering the common support. We also evaluate the computational complexity of the \ac{GSOMP} to showcase its efficiency with respect to the \ac{OMP}. Numerical results show that the propounded estimator outperforms the \ac{OMP}-based estimator in the low and moderate \ac{SNR} regimes, whilst achieving the same accuracy in the high \ac{SNR} regime.

\item We analyze the mean-square error of the \ac{GSOMP} scheme by providing the \ac{CRLB}. Moreover, we calculate the average achievable rate assuming imperfect channel gain knowledge at the \ac{BS}. We then show numerically that when the angle quantization error involved in the sparse channel representation is negligible, the performance of the \ac{GSOMP}-based estimator is very close to the \ac{CRLB}. Additionally, the average achievable rate approaches that of the perfect channel knowledge case at moderate and high \ac{SNR} values, hence corroborating the good performance of our design. Finally, we extend our analysis to the case of a multi-antenna user, and discuss the benefits of deploying multiple antennas at the user side.
\end{itemize}
The rest of this paper is organized as follows: Section~\ref{sec:system model} introduces the system and channel models. Section~\ref{sec:hybrid_bf} describes the hybrid combining problem under the beam squint effect, and presents the proposed combining scheme. Section~\ref{sec:channel_estimation} formulates the channel estimation problem, introduces the standard estimation methods, and explains the propounded algorithm for estimating the \ac{SFW} channel. Section~\ref{sec:multiantenna_users} extends the analysis to the multi-antenna user case. Section~\ref{sec:numerical_results} is devoted to numerical simulations. Finally, Section~\ref{sec:conclusions} summarizes the main conclusions derived in this work.

\textit{Notation}: Throughout the paper, $D_{N}(x) = \frac{\sin(Nx/2)}{N\sin(x/2)}$ is the Dirichlet sinc function; $\mathbf{A}$ is a matrix; $\mathbf{a}$ is a vector; $a$ is a scalar; $\mathbf{A}^{\dagger}$, $\mathbf{A}^{H}$, and $\mathbf{A}^{T}$ are the pseudoinverse, conjugate transpose, and transpose of $\mathbf{A}$, respectively; $\mathbf{A}(i)$ is the $i$th column of matrix $\mathbf{A}$; $\mathbf{A}(\mathcal{I})$ is the submatrix containing the columns of $\mathbf{A}$ given by the indices set $\mathcal{I}$; $|\mathcal{I}|$ is the cardinality of set $\mathcal{I}$; $\text{tr}\{\mathbf{A}\}$ is the trace of $\mathbf{A}$; $\text{blkdiag}(\mathbf{A}_1,\dots, \mathbf{A}_n)$ is the block diagonal matrix; $[\mathbf{A}]_{n,m}$ is the $(n,m)$th element of matrix $\mathbf{A}$; $\mathcal{F}\{\cdot\}$ denotes the continuous-time Fourier transform; $\ast$ denotes convolution; $\text{Re}\{\cdot\}$ is the real part of a complex variable; $\mathbf{1}_{N\times M}$ is the $N\times M$ matrix with unit entries; $\mathbf{I}_{N}$ is the $N\times N$ identity matrix; $[\mathbf{v}]_n$ is the $n$th entry of vector $\mathbf{v}$; $\text{supp}(\mathbf{v}) = \{n: [\mathbf{v}]_n \neq 0\}$ is the support of $\mathbf{v}$; $\otimes$ denotes the Kronecker product; $\odot$ is the element-wise product; $\|\mathbf{a}\|_1$ and $\|\mathbf{a}\|_2$ are the $l_1$-norm and $l_2$-norm of vector $\mathbf{a}$, respectively; $\delta(\cdot)$ is the Kronecker delta function; $\mathbb{E}\{\cdot\}$ denotes expectation; and $\mathcal{CN}(\bm{\mu}, \mathbf{R})$ is a complex Gaussian vector with mean $\bm{\mu}$ and covariance matrix $\mathbf{R}$.
\begin{figure*}
	\centering
	\begin{subfigure}{.5\textwidth}
		\centering
		\includegraphics[width=.64\linewidth]{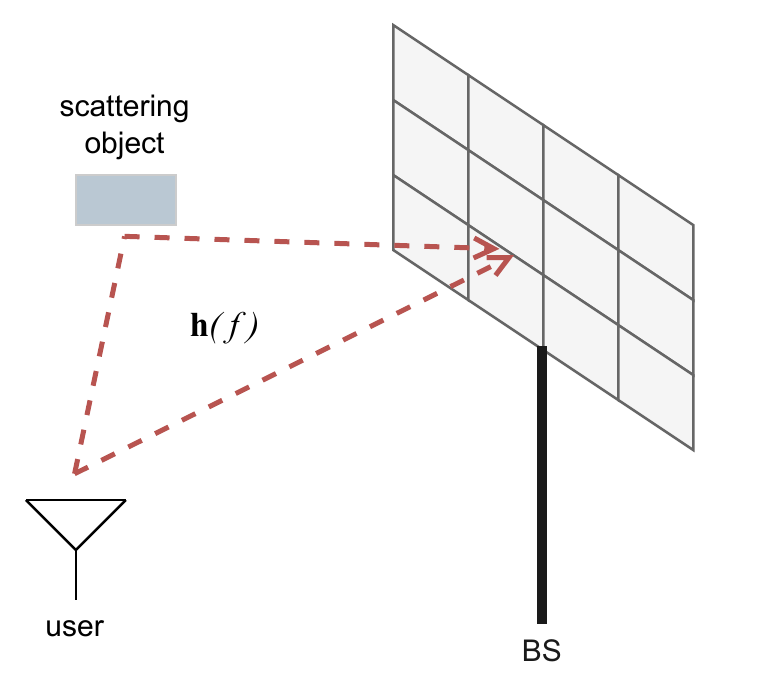}
		\caption{\footnotesize Uplink setup}
		\label{fig:Fig_sm_us}
	\end{subfigure}%
	\begin{subfigure}{.5\textwidth}
		\centering
		\includegraphics[width=.76\linewidth]{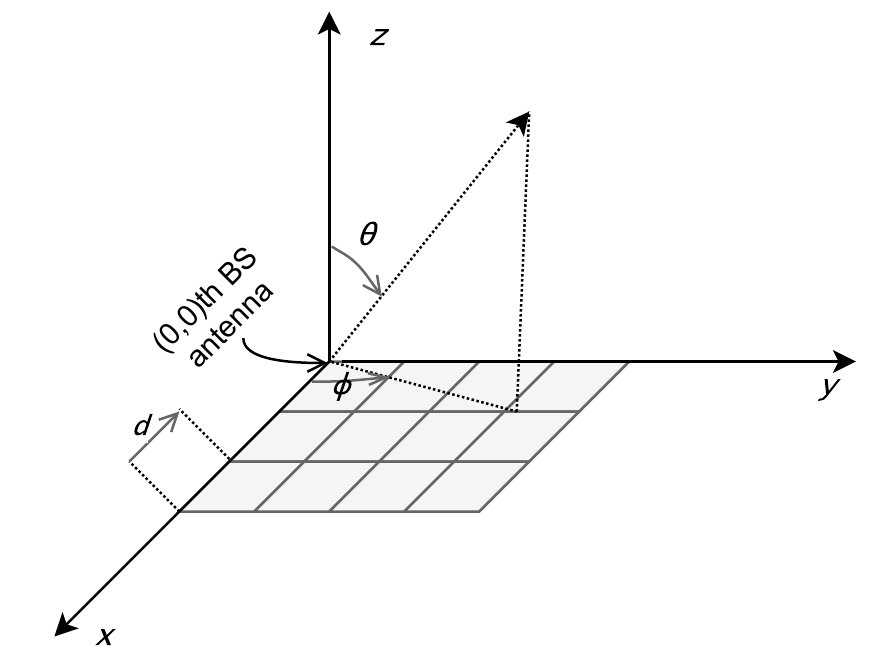}
		\caption{\footnotesize Array geometry}
		\label{fig:Fig_sm_ag}
	\end{subfigure}%
	\caption{Illustration of the \ac{BS} antenna array and its geometry considered in the system model.}
	\label{fig:Fig_sm}
\end{figure*}
\begin{table}[H]
	\centering
	\caption{Main Notation Used in the System Model}
	\label{Table:Notation}
	\begin{tabular}{|l  l|}
		\hline
		\textbf{Notation} & \textbf{Description} \\
		\hline
		$N_B$ & Number of \ac{BS} antennas \\
		$N_{\text{RF}}$ & Number of \ac{RF} chains \\
		\hline
		$S$ & Number of subcarriers\\
		$f_s$ & Frequency of the $s$th subcarrier \\
		$ B $ & Total signal bandwidth\\
		\hline
		$L$ & Number of NLoS paths\\
		$\alpha_l(f)$ & Frequency-selective attenuation of the $l$th path\\
		$\tau_l$ & ToA of the $l$th path\\
		$(\phi_l,\theta_l)$ & DoA of the $l$th path\\
		$\tau_{l,nm}$ & Time delay to the $(n,m)$th BS antenna over the $l$th path\\
		$\tau_{nm}(\phi_l,\theta_l)$ & Time delay from the $(0,0)$th to the $(n,m)$th BS antenna \\
		\hline
		$\mathtt{x}(t)$ & Baseband-equivalent of transmitted signal\\
		$x(f)$ & Fourier transform of $\mathtt{x}(t)$\\
		$\mathtt{x}_l(t)$ & Distorted version of $\mathtt{x}(t)$  over the $l$th path\\
		\hline
		$\tilde{r}_{nm}(t)$ & Passband signal received by the $(n,m)$th BS antenna \\
		$\mathtt{r}_{nm}(t)$ & Baseband-equivalent of $\tilde{r}_{nm}(t)$\\
		$r_{nm}(f)$ & Fourier transform of $\mathtt{r}_{nm}(t)$\\
		\hline
		$d$ & Antenna spacing\\
		$f_c$ & Carrier frequency \\
		$c$ & Speed of light \\
		\hline
		$k_{\text{abs}}$ & Molecular absorption coefficient\\
		$\mathtt{D}$ & Distance between the \ac{BS} and the user \\
		$\Gamma_l(f)$ & Reflection coefficient of the $l$th NLoS path \\
		\hline
	\end{tabular}
\end{table}
\section{System Model}\label{sec:system model}
Consider the uplink of a \ac{THz} massive \ac{MIMO} system where the \ac{BS} is equipped with an $N\times M$-element \ac{UPA}, and serves a single-antenna user as depicted in Fig~\ref{fig:Fig_sm}(\subref{fig:Fig_sm_us}); the multi-antenna user case is investigated in Section~\ref{sec:multiantenna_users}. The total number of \ac{BS} antennas is $N_B = NM$, and the baseband frequency response of the uplink channel is denoted by $\mathbf{h}(f)\in\mathbb{C}^{N_B\times 1}$. In the sequel, we present the channel and hybrid transceiver models used in this work.

\subsection{\ac{THz} Channel Model with Spatial-Wideband Effects}
Due to limited scattering in THz bands, the propagation channel is represented by a ray-based model of $L+1$ rays~\cite{THz_spatial_modulation},~\cite{THz_Prop_models}. Hereafter, we assume that the $0$th ray corresponds to the \ac{LoS} path, while the remaining $l=1,\dots,L$, rays are \ac{NLoS} paths. Specifically, each path $l=0,\dots, L$, is characterized by its frequency-selective path attenuation $\alpha_l(f)$, \ac{ToA} $\tau_l$, and \ac{DoA} $(\phi_l,\theta_l)$, where $\phi_l\in[-\pi,\pi]$ and $\theta_l\in[-\frac{\pi}{2},\frac{\pi}{2}]$ are the azimuth and polar angles, respectively. In the far-field region\footnote{Near-field considerations are provided in Section~\ref{sec:spherical_wave}.} of the \ac{BS} antenna array, the total delay between the user and the $(n,m)$th BS antenna through the $l$th path, $\tau_{l,nm}$, is calculated as
\begin{equation}{\label{eq:totaldelay}}
\tau_{l,nm} = \tau_l + \tau_{nm}(\phi_l,\theta_l),
\end{equation}
where $\tau_{nm}(\phi_l,\theta_l)$ accounts for the propagation delay across the BS array, and is measured with respect to the $(0,0)$th \ac{BS} antenna. For a UPA placed in the $xy$-plane (see Fig.~\ref{fig:Fig_sm}(\subref{fig:Fig_sm_ag})), we then have~\cite{antenna_th1}
\begin{equation}\label{eq:upa_timedelay}
\tau_{nm}(\phi_l,\theta_l) \triangleq \frac{d(n \sin\theta_l\cos\phi_l+ m\sin\theta_l\sin\phi_l)}{c},
\end{equation}
where $d$ is the antenna separation, and $c$ is the speed of light. The channel frequency response is derived as follows. Let $\mathtt{x}(t)$ be the baseband signal transmitted by the user, with $\mathcal{F}\{\mathtt{x}(t)\} = x(f)$. The passband signal, $\tilde{r}_{nm}(t)$, received by the $(n,m)$th BS antenna is written in the noiseless case as~\cite{tse_book}
\begin{equation}
\tilde{r}_{nm}(t) = \sum_{l=0}^L \sqrt{2}\text{Re}\left\{\mathtt{x}_l(t-\tau_{l,nm})e^{j2\pi f_c (t-\tau_{l,nm})}\right\},
\end{equation}
where $f_c$ is the carrier frequency, $\mathtt{x}_{l}(t) \triangleq \mathtt{x}(t)\ast \chi_l(t)$ is the distorted baseband waveform due to the frequency-selective attenuation of the $l$th path, and $\chi_l(t)$ models the said distortion; namely, $\mathcal{F}\{\chi_l(t)\} = \alpha_l(f)$ and $\mathcal{F}\{\mathtt{x}_{l}(t)\} = \alpha_l(f)x(f)$~\cite{uwb_paper}. Next, the received passband signal $\tilde{r}_{nm}(t)$ is down-converted to the baseband signal $\mathtt{r}_{nm}(t)$, which is given by 
\begin{equation}\label{eq:baseband_signal}
\mathtt{r}_{nm}(t) =  \sum_{l=0}^L e^{-j2\pi f_c\tau_l} e^{-j2\pi f_c \tau_{nm}(\phi_l,\theta_l) }\mathtt{x}_l(t-\tau_{l,nm}).
\end{equation}
Taking the continuous-time Fourier transform of~\eqref{eq:baseband_signal} yields
\begin{align}
r_{nm}(f) &= \mathcal{F}\{\mathtt{r}_{nm}(t)\} \nonumber\\
&=\sum_{l=0}^{L} \beta_l(f) e^{-j2\pi (f_c+f)\tau_{nm}(\phi_l,\theta_l)} x(f) e^{-j2\pi f\tau_l},
\end{align}
where $\beta_l(f) \triangleq \alpha_l(f)e^{-j2\pi f_c\tau_l}$ is the complex gain of the $l$th path. Lastly, collecting all $r_{nm}(f)$ into a vector $\mathbf{r}(f)\in~\mathbb{C}^{N_B\times 1}$ gives the relation $\mathbf{r}(f)= \mathbf{h}(f)x(f)$, where
\begin{align}\label{eq:physical_model}
\mathbf{h}(f) = \sum_{l=0}^{L}\beta_l(f) \mathbf{a}(\phi_l,\theta_l, f)e^{-j2\pi f\tau_l}
\end{align}
is the baseband frequency response of the uplink channel, and 
\begin{multline}\label{eq:upa_response}
\mathbf{a}(\phi,\theta,f)= \left[ 1, \dots,e^{-j2\pi (f_c+f)\frac{d}{c}(n \sin\theta\cos\phi+ m\sin\theta\sin\phi)}\right .,\\ 
\left . \dots, e^{-j2\pi (f_c+f)\frac{d}{c}((N-1) \sin\theta\cos\phi+ (M-1)\sin\theta\sin\phi)}\right]^T
\end{multline}
is the array response vector of the \ac{BS}. Here, the array response is frequency-dependent due to the \textit{spatial-wideband} effect.\footnote{If the delay across the BS array is small relative to the symbol period, then $\mathtt{x}_l(t-\tau_{l,nm})\approx\mathtt{x}_l(t-\tau_{l})$. In this case, we have a spatially narrowband channel with frequency-flat array response vector, i.e., $\mathbf{a}(\phi,\theta,0)$.}

We now introduce the path attenuation model. First, the so-called molecular absorption loss is no longer negligible at \ac{THz} frequencies. Therefore, the path attenuation of the \ac{LoS} path is calculated as~\cite{analytical_perf_thz}
\begin{equation}\label{eq:path_atten_LoS}
|\beta_0(f)| = \alpha_0(f) = \frac{c}{4\pi (f_c + f) \mathtt{D}} e^{-\frac{1}{2}k_{\text{abs}}(f_c+ f) \mathtt{D}},
\end{equation}
where $\mathtt{D}$ denotes the distance between the \ac{BS} and the user, and $k_{\text{abs}}(\cdot)$ is the molecular absorption coefficient determined by the composition of the propagation medium; different from \ac{mmWave} channels, the major molecular absorption in \ac{THz} bands comes from water vapor molecules~\cite{analytical_perf_thz}. For the \ac{NLoS} paths, we consider single-bounce reflected rays, since the diffused and diffracted rays are heavily attenuated for distances larger than a few meters~\cite{multiray_modeling_thz}. To this end, the reflection coefficient, $\Gamma_l(f)$, should be taken into account, which is specified~as~\cite{scattering_thz}
\begin{equation} 
\Gamma_l(f) = \frac{\cos\phi_{i,l} - n_t\cos\phi_{t,l}}{\cos\phi_{i,l} + n_t\cos\phi_{t,l}} e^{-\left(\frac{8\pi^2 (f_c+f)^2\sigma^2_{\text{rough}}\cos^2\phi_{i,l}}{c^2}\right)},
\end{equation}	
where $n_t\triangleq Z_0/Z$ is the refractive index, $Z_0=377$ Ohm is the free-space impedance, $Z$ is the impedance of the reflecting material, $\phi_{i,l}$ is the incidence and reflection angle, $\phi_{t,l}= \text{arcsin}\left(n^{-1}_t\sin\phi_{i,l}\right)$ is the refraction angle, and $\sigma_{\text{rough}}$ characterizes the roughness of the reflecting surface. The path attenuation of the $l$th \ac{NLoS} path is finally given by~\cite{nlos_pl} 
\begin{equation}\label{eq:path_atten_NLoS}
|\beta_l(f)| =  \alpha_{l}(f) = |\Gamma_l(f)|\alpha_{0}(f),
\end{equation} 
where $l=1,\dots, L$.

\subsection{Hybrid Transceiver Model}
\begin{figure}[t]
	\centering	
	\includegraphics[width=92mm]{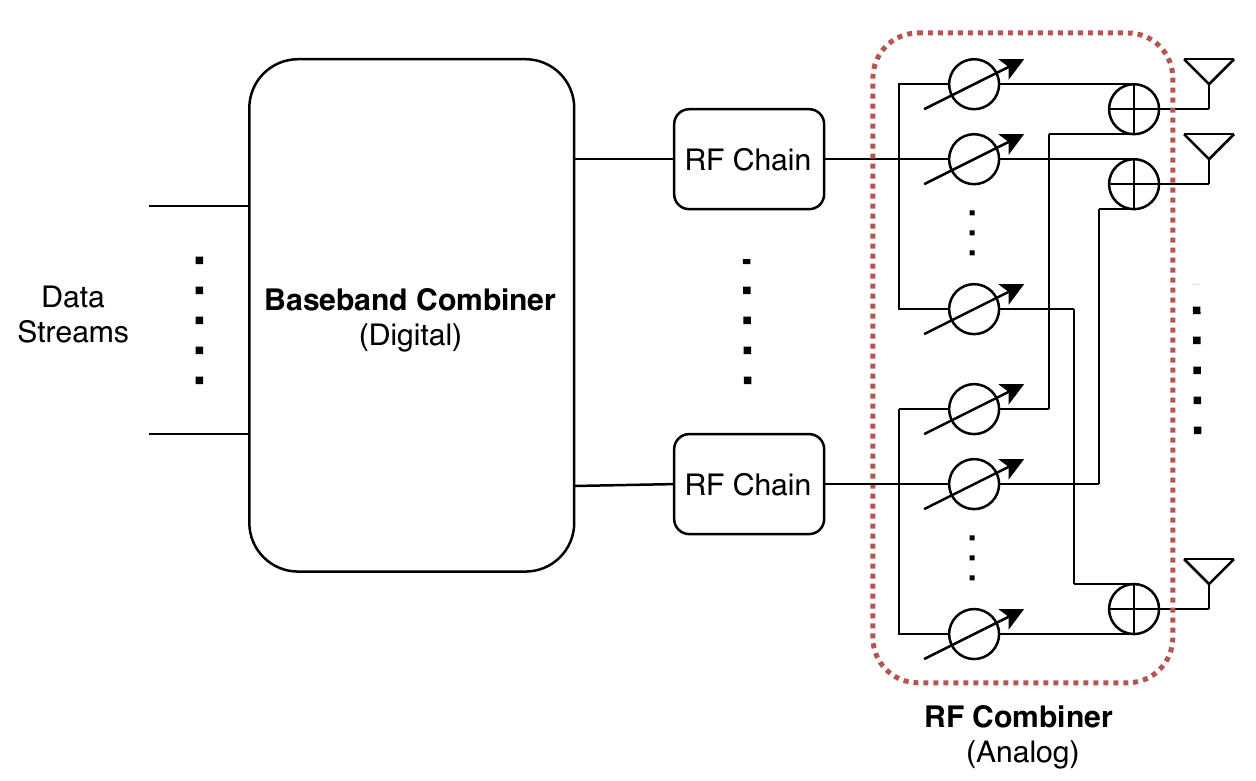}
	\caption{Illustration of the hybrid array structure considered in the system model.}
	\label{fig:Fig0}
\end{figure}

Due to the frequency selectivity of the \ac{THz} channel, \ac{OFDM} modulation is employed to combat inter-symbol interference. Specifically, we consider $S$ subcarriers over a signal bandwidth~$B$. Then, the baseband frequency of the $s$th subcarrier is specified as $f_s = \left(s - \frac{S-1}{2}\right)\frac{B}{S}, s=0,\dots, S-1$. A hybrid analog-digital architecture with $N_{\text{RF}}\ll N_B$ \ac{RF} chains is also considered at the \ac{BS} to facilitate efficient hardware implementation; each \ac{RF} chain drives the array through $N_B$ analog phase shifters, as shown in Fig.~\ref{fig:Fig0}. The hybrid combiner for the $s$th subcarrier is hence expressed as $\mathbf{F}[s] = \mathbf{F}_{\text{RF}}\mathbf{F}_{\text{BB}}[s] \in\mathbb{C}^{N_B\times N_{\text{RF}}}$, where $\mathbf{F}_{\text{RF}}\in\mathbb{C}^{N_B\times N_{\text{RF}}}$ is the frequency-flat \ac{RF} combiner with elements of constant amplitude, i.e., $\frac{1}{\sqrt{N_B}}$, but variable phase, and $\mathbf{F}_{\text{BB}}[s]\in\mathbb{C}^{N_{\text{RF}}\times N_{\text{RF}}}$ is the baseband combiner. Finally, the post-processed baseband signal, $\mathbf{y}[s]\in\mathbb{C}^{N_{\text{RF}}\times 1}$, for the $s$th subcarrier is written as
\begin{align}\label{eq:post_processed}
\mathbf{y}[s] & = \mathbf{F}^H[s]\mathbf{r}[s] \nonumber \\
&=\mathbf{F}^H[s]\left(\sqrt{P_d}\mathbf{h}[s]x[s] + \mathbf{n}[s]\right),
\end{align}
where $\mathbf{r}[s] \triangleq \mathbf{r}(f_s)$ and $\mathbf{h}[s]\triangleq \mathbf{h}(f_s)$ are the received signal and uplink channel, respectively, $x[s]\triangleq x(f_s) \sim\mathcal{CN}(0,1)$ is the data symbol transmitted at the $s$th subcarrier, $P_d$ denotes the average power per data subcarrier assuming equal power allocation among subcarriers, and $ \mathbf{n}[s]\sim\mathcal{CN}(\mathbf{0}, \sigma^2\mathbf{I}_{N_{B}})$ is the additive noise vector. 

\begin{remark}
A promising alternative to \ac{OFDM} is single-carrier with frequency domain equalization (SC-FDE) due to its favorable \ac{PAPR}. In our work, we exploit the inherent characteristics of \ac{THz} channels, i.e., high path loss and directional transmissions, which result in a coherence bandwidth of hundreds of MHz~\cite{multiray_modeling_thz}. Therefore, a relatively small number of subcarriers is used, which is expected to yield a tolerant \ac{PAPR}.  
\end{remark}

\section{Hybrid Combining}\label{sec:hybrid_bf}

\subsection{The Beam Squint Problem}
Even for a moderate number of BS antennas, the propagation delay across the array can exceed the sampling period due to the ultra-high bandwidth used in THz communication. As a result, the \ac{DoA}/\ac{DoD} varies across the \ac{OFDM} subcarriers, and the array gain becomes frequency-selective. This phenomenon, known as \textit{beam squint} in the array processing literature, calls for a frequency-dependent combining design which is feasible only in a fully-digital array architecture. 

\begin{figure*}[t]
	\centering
	\includegraphics[width=1\linewidth]{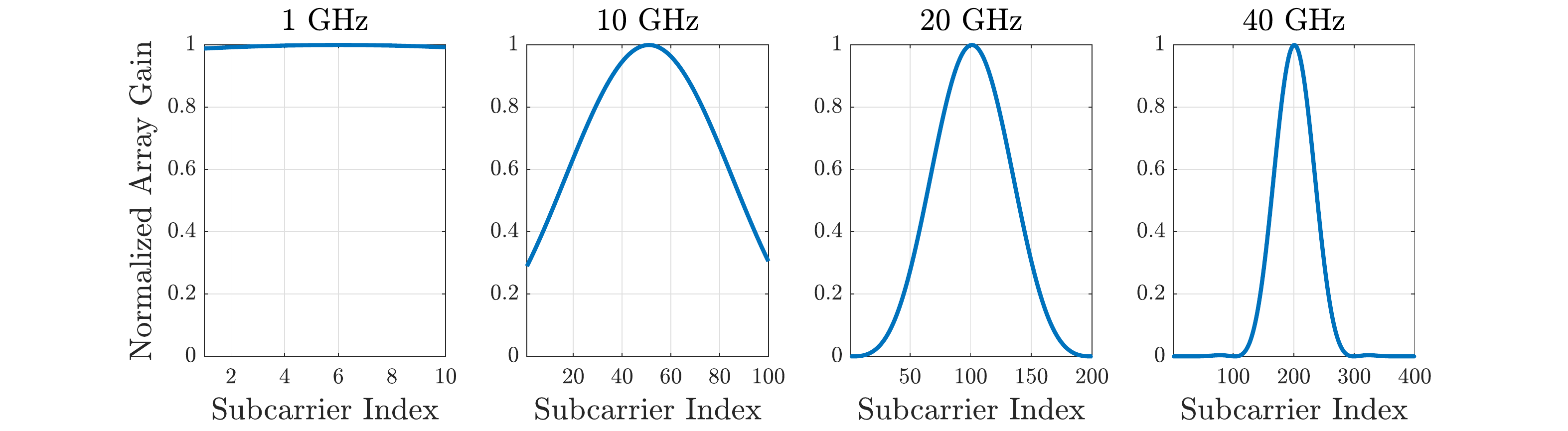}
	\caption{Normalized array gain for various bandwidths; $100\times 100$-element UPA, $f_c = 300$ GHz, coherence bandwidth of $100$ MHz, and $(\phi,\theta) = (\pi/3, \pi/4)$.}
	\label{fig:Fig1}
\end{figure*}

To demonstrate the detrimental effect of beam squint when frequency-flat \ac{RF} combining is employed, we consider an arbitrary ray impinging on the BS array with \ac{DoA} $(\phi,\theta)$; therefore, we omit the subscript ``$l$' hereafter. In the narrowband case, the uplink channel is described as $\beta\mathbf{a}(\phi,\theta,0)$. Let $\mathbf{f}_{\text{RF}} = (1/\sqrt{N_B})\mathbf{f}$, with $\|\mathbf{f}\|^2 = N_B$, be an arbitrary \ac{RF} combiner. For the combiner $\mathbf{f}_{\text{RF}}$, the power of the received signal is calculated as
\begin{equation}\label{eq:rx_power_singleantenna}
|\beta|^2 \frac{\left|\mathbf{f}^H\mathbf{a}(\phi,\theta,0)\right|^2}{N_B}P_d = |\beta|^2 N_BG(\phi,\theta,0)P_d,
\end{equation}
where $G(\phi,\theta,f) \triangleq |\mathbf{f}^H\mathbf{a}(\phi,\theta,f)|^2 /N_B^2$ is the \textit{normalized array gain}. Choosing $\mathbf{f} = \mathbf{a}(\phi,\theta,0)$ yields $G(\phi,\theta,0) =1$, and the maximum array gain is obtained. In a wideband THz system, though, the array gain varies across the OFDM subcarriers. In particular, we have that 
\begin{align}\label{eq:array_gain}
G(\phi,\theta,f) &= \frac{|\mathbf{a}^H(\phi,\theta,0)\mathbf{a}(\phi,\theta,f)|^2}{N_B^2}\nonumber\\
&= \left|D_{N}(2\pi f\Delta_x(\phi,\theta))\right|^2  \left | D_{M}(2\pi f\Delta_y(\phi,\theta)) \right|^2,
\end{align}
where $\Delta_x(\phi,\theta) \triangleq (d\sin\theta\cos\phi)/c$ and $\Delta_y(\phi,\theta) \triangleq (d\sin\theta\sin\phi)/c$; please refer to Appendix A for the proof. Figure~\ref{fig:Fig1} shows the array gain for various bandwidths, when the narrowband \ac{RF} combiner $\mathbf{f}_{\text{RF}} = 1/(\sqrt{N_B})\mathbf{a}(\phi,\theta,0)$ is used. As we see, the array gain reduces substantially across the OFDM subcarriers. Furthermore, using the technique of~\cite{optimality_beam_steering}, one can show that $G(\phi,\theta,f)\to 0$ as $NM\to\infty$. 
Contrary to narrowband massive MIMO, where the signal power increases monotonically with the number of BS antennas, here it may decrease. Consequently, beam squint compensation is of paramount importance for the successful deployment of THz massive MIMO systems. 

\subsection{Proposed Combiner for Single-Path Channels}
In this section, we introduce our wideband combining scheme for single-path channels, and then extend it to the multi-path case. To this end, we consider that the BS employs a single \ac{RF} chain to combine the incoming signal, and hence the \ac{RF} combiner is denoted by $\mathbf{f}_{\text{RF}}$. Next, we analyze the normalized array gain by decomposing the array into $N_{\text{sb}}\times M_{\text{sb}}$ virtual subarrays of $\tilde{N}\tilde{M}$ antennas each, where $\tilde{N} \triangleq N/N_{\text{sb}}$ and $\tilde{M} \triangleq M/M_{\text{sb}}$. 

\subsubsection{Virtual Array Partition} The array response vector in~\eqref{eq:upa_response} is decomposed as
\begin{equation}\label{eq:upa_response_vector}
\mathbf{a}(\phi,\theta,f) = \mathbf{a}_x(\phi,\theta,f) \otimes \mathbf{a}_y(\phi,\theta,f),
\end{equation}
where $\mathbf{a}_x(\cdot)$ and $\mathbf{a}_y(\cdot)$ are defined, respectively, as
\begin{multline}\label{eq:ax_upa}
\mathbf{a}_x(\phi,\theta,f) \triangleq \left[1, \dots,e^{-j2\pi (f_c+f)n\Delta_x(\phi,\theta)},\right . \\
\left . \dots, e^{-j2\pi (f_c+f)(N-1)\Delta_x(\phi,\theta)}\right]^T
\end{multline}
and
\begin{multline}\label{eq:ay_upa}
\mathbf{a}_y(\phi,\theta,f) \triangleq \left[1, \dots,e^{-j2\pi (f_c+f)m\Delta_y(\phi,\theta)},\right .\\
\left . \dots, e^{-j2\pi (f_c+f)(M-1)\Delta_y(\phi,\theta)}\right]^T.
\end{multline}
Using the previously mentioned virtual array partition, we can write
\begin{align}
\mathbf{a}_x(\phi,\theta, f) &= \left[\mathbf{a}_{x,1}(\phi,\theta,f),\dots, \mathbf{a}_{x,N_{\text{sb}}}(\phi,\theta,f)\right]^T,\\
\mathbf{a}_y(\phi,\theta, f) &=  \left[\mathbf{a}_{y,1}(\phi,\theta,f),\dots, \mathbf{a}_{y,M_{\text{sb}}}(\phi,\theta,f)\right]^T,
\end{align}
where $\mathbf{a}_{x,n}(\phi,\theta,f)$ corresponds to the response vector of the $n$th virtual subarray, and is defined as
\begin{multline}
\mathbf{a}_{x,n}(\phi,\theta,f)\triangleq \left[e^{-j2\pi(f_c + f)(n-1)\tilde{N}\Delta_x(\phi,\theta)},\right. \\
\left . \dots, e^{-j2\pi(f_c + f)(n\tilde{N}-1)\Delta_x(\phi,\theta)}\right]^T.
\end{multline}
Finally, each vector $\mathbf{a}_{x,n}(\phi,\theta,f)$ is expressed in terms of $\mathbf{a}_{x,1}(\phi,\theta,f)$, i.e., the response of the \textit{first subarray}, as 
\begin{equation}
\mathbf{a}_{x,n}(\phi, \theta, f) = e^{-j2\pi(f_c + f)(n-1)\tilde{N}\Delta_x(\phi,\theta)}\mathbf{a}_{x,1}(\phi, \theta,f).
\end{equation}
\begin{figure*}[b]
	\hrulefill
	\begin{align}\label{eq:array_gain_virtualpart}
	G(\phi,\theta,f) &= \frac{\left |\mathbf{a}^H_{x,1}(\phi, \theta, 0)\mathbf{a}_{x,1}(\phi, \theta, f)\right |^2\left |\mathbf{a}^H_{y,1}(\phi, \theta, 0)\mathbf{a}_{y,1}(\phi, \theta, f)\right |^2}{\tilde{N}^2\tilde{M}^2}
	\underbrace{\frac{\left |\sum_{n=1}^{N_{\text{sb}}} \sum_{m=1}^{M_{\text{sb}}}e^{-j2\pi (n-1) \tilde{N} f \Delta_x(\phi,\theta)} e^{-j2\pi (m-1) \tilde{M} f \Delta_y(\phi,\theta)} \right |^2}{N^2_{\text{sb}}M^2_{\text{sb}}}}_{\Omega(\phi,\theta,f)} \nonumber\\
	& = \left |D_{\tilde{N}}(2\pi f\Delta_x(\phi,\theta))\right |^2\left |D_{\tilde{M}}(2\pi f\Delta_y(\phi,\theta))\right |^2\Omega(\phi,\theta, f).
	\end{align}
\end{figure*}
We stress that similar expressions hold for the vector $\mathbf{a}_y$. Using the virtual subarray notation, the normalized array gain $G(\phi,\theta, f)$ is recast as in~\eqref{eq:array_gain_virtualpart} at the bottom of the next page. For an adequately small $\tilde{N}\tilde{M}$, we then have the approximation $D_{\tilde{N}}(2\pi f_s\Delta_x(\phi,\theta))D_{\tilde{M}}(2\pi f_s\Delta_y(\phi,\theta)) \approx 1$. 
\begin{figure}[H]
	\centering	
	\includegraphics[width=58mm]{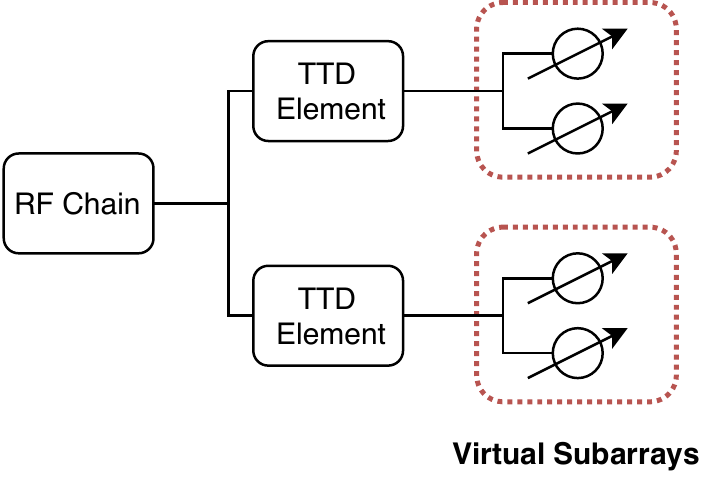}
	\caption{Illustration of the TTD-based wideband combiner with virtual array partition; the circles with arrows represent frequency-flat phase shifters.}
	\label{fig:Fig00}
\end{figure} 

\subsubsection{Size of Virtual Subarrays}
The size of each virtual subarray, $\tilde{N}\times \tilde{M}$, is selected such that the maximum delay across the first virtual subarray is smaller than the sampling period $1/B$. Specifically, the maximum delay, $\tau_{\max}$, across the first subarray is given by~\eqref{eq:upa_timedelay} for $n=\tilde{N}-1$, $m=\tilde{M}-1$, $\sin\theta = 1$, and $\sin\phi = \cos\phi = 1/\sqrt{2}$, yielding $\tau_{\max} = d(\tilde{N}+\tilde{M}-2)/(\sqrt{2}c)$. For half-wavelength antenna spacing and $\tilde{N}=\tilde{M}$, the condition $\tau_{\max} < 1/B$ reduces to $(\tilde{N} - 1) < \sqrt{2}f_c/B$, which is used to determine $\tilde{N}$.

\subsubsection{\ac{TTD}-Based Combining} The factor $\Omega(\phi,\theta, f)\leq 1$ in~\eqref{eq:array_gain_virtualpart} accounts for the losses caused by the delay between consecutive virtual subarrays, and it can be canceled through a \ac{TTD} network placed between virtual subarrays, as depicted in Fig.~\ref{fig:Fig00}. Then, we obtain $\Omega(\phi,\theta, f_s) = 1$ by multiplying the signal at the $(n,m)$th virtual subarray by $e^{j2\pi f_s\Delta_{mn}(\phi,\theta)}$, where $\Delta_{mn}(\phi,\theta)\triangleq(n-1)\tilde{N} \Delta_x(\phi,\theta) + (m-1)\tilde{M} \Delta_y(\phi,\theta)$ is the delay to be mitigated. Because all \ac{OFDM} subcarriers share the same delay $\Delta_{mn}(\phi,\theta)$, it can be compensated using a single \ac{TTD} element modeled as a linear filter with impulse response $\delta(t - \Delta_{nm}(\phi,\theta))$. Therefore, the wideband RF combiner is designed as
\\
\begin{equation}\label{eq:ttd_bf_upa}
\mathbf{f}_{\text{RF}}[s] = \frac{1}{\sqrt{N_B}}\text{vec}\left(\mathbf{A}(\phi,\theta, 0) \odot \mathbf{T}[s]\right),
\end{equation}
\\
where $\mathbf{T}[s] \triangleq \left[e^{-j2\pi f_s\Delta_{mn}(\phi,\theta)}\right]_{m=1,n=1}^{M_{\text{sb}},N_{\text{sb}}}\otimes \mathbf{1}_{\tilde{M}\times \tilde{N}}$, $\mathbf{A}(\phi,\theta, 0)\triangleq  \mathbf{a}_y(\phi,\theta,0)\mathbf{a}^T_x(\phi,\theta,0)$, and $\|\mathbf{f}_{\text{RF}}[s]\|^2 = 1$. 
\begin{proposition}
With the proposed combiner \eqref{eq:ttd_bf_upa}, we have
\begin{equation}\label{eq:array_gain_proposition}
\left |\mathbf{f}^H_{\emph{RF}}\mathbf{a}(\phi, \theta, f)\right |^2 = N_B \left |D_{\tilde{N}}(2\pi f\Delta_x)\right |^2 \left | D_{\tilde{M}}(2\pi f\Delta_y)\right |^2,
\end{equation}
where $D_{N}(x) = \frac{\sin(Nx/2)}{N\sin(x/2)}$ is the Dirichlet sinc function.
\begin{proof}
See Appendix B.
\end{proof}	
\end{proposition}
 From~\eqref{eq:array_gain_proposition}, we conclude that for sufficiently small $\tilde{N}$ and $\tilde{M}$, an array gain $N_B$ is approximately achieved over the whole signal bandwidth $B$. Thus, the SNR at the $s$th OFDM subcarrier is $|\beta(f_s)|^2N_B P_d/\sigma^2$. Lastly, $(N_{\text{sb}}M_{\text{sb}}-1)$ TTD elements are employed per RF chain, where $N_{\text{sb}} = N/\tilde{N}$ and $M_{\text{sb}} = M/\tilde{M}$.
  
\subsection{Proposed Combiner for Multi-Path Channels} 
The propounded method can readily be applied to multi-path channels. For example, consider a THz channel comprising of $L=2$ \ac{NLoS} paths. In a fully-digital array, the optimal combiner for the $s$th subcarrier is given by the maximum-ratio combiner $\mathbf{h}[s]/\|\mathbf{h}[s]\|$. By employing $N_{\text{RF}}=2$ RF chains, we have that 
\begin{equation}
\frac{\mathbf{h}[s]}{\|\mathbf{h}[s]\|} = \mathbf{F}_{\text{RF}}[s] \mathbf{F}_{\text{BB}}[s] \mathbf{1}_{2\times 1},
\end{equation}
where 
\begin{align}\label{eq:opt_pre_widemultipath}
\mathbf{F}_{\text{RF}}[s] & = \frac{1}{\sqrt{N_B}}
\begin{bmatrix}
\mathbf{a}\left(\phi_1, \theta_1, f_s\right) & \mathbf{a}\left(\phi_2, \theta_2,f_s\right)
\end{bmatrix}, \\[0.2cm]
\mathbf{F}_{\text{BB}}[s] &= \frac{\sqrt{N_B}}{|\mathbf{h}[s]|}
\begin{bmatrix}
\beta_1(f_s)e^{-j2\pi f_s \tau_1} & 0 \\
0 & \beta_2(f_s)e^{-j2\pi f_s \tau_2}
\end{bmatrix}.
\end{align}
The columns of the \textit{wideband} \ac{RF} combiner $\mathbf{F}_{\text{RF}}[s]$ are then approximated using~\eqref{eq:ttd_bf_upa}, whilst the vector $\mathbf{1}_{2\times 1}$ with unit entries performs the addition of the two outputs of the baseband combiner. Note that $N_{\text{RF}} = L$ are required to implement the maximum-ratio combiner in a hybrid array architecture. 

\begin{remark}	
A few recent papers in the literature (e.g.,~\cite{wideband_multibeam} and references therein) suggested the use of \ac{TTD} to provide frequency-dependent phase shifts at each antenna of an $N\text{-element}$ ULA, yielding a wideband multi-beam architecture. In our work, we adopt a hybrid array architecture, where each frequency-independent phase shifter drives a single antenna whilst each \ac{TTD} element controls a group of antennas, i.e., virtual subarray. Moreover, we consider a \ac{UPA}, and hence our design enables squint-free \ac{3D} combining.
\end{remark}

\section{Sparse Channel Estimation}\label{sec:channel_estimation}
We have introduced an effective wideband combiner assuming that the BS has perfect knowledge of the uplink channel. In this section, we investigate the channel estimation problem under the \textit{spatial-wideband} effect. More particularly, we first formulate a compressive sensing problem to estimate the channel at each subcarrier independently with reduced training overhead. We then propound a wideband dictionary and employ an estimation algorithm that leverages information from multiple subcarriers to increase the reliability of the channel estimates in the low and moderate \ac{SNR} regimes. 

\subsection{Problem Formulation}
We assume a block-fading model where the channel coherence time is much larger than the training period. Specifically, the training period consists of $N_{\text{slot}}$ time slots. At each time slot $t= 1,\dots, N_{\text{slot}}$, the user transmits the pilot signal $x_t[s] = \sqrt{P_p}, \ \forall s\in\mathcal{S}$, where $\mathcal{S} \triangleq \{1,\dots, S\}$ denotes the set of \ac{OFDM} subcarriers, and $P_p$ is the power per pilot subcarrier. In turn, the BS combines the pilot signal at each subcarrier~$s\in\mathcal{S}$ using a training hybrid combiner $\mathbf{W}_t[s]\in\mathbb{C}^{N_B\times N_{\text{RF}}}$. Therefore, the post-processed signal at slot $t$, $\mathbf{y}_t[s]\in\mathbb{C}^{N_{\text{RF}}\times 1}$, is written as
\begin{equation}
\mathbf{y}_t[s] = \sqrt{P_p} \mathbf{W}_t^H[s]\mathbf{h}[s] + \mathbf{W}_t^H[s]\mathbf{n}_t[s],
\end{equation}
where $\mathbf{n}_t[s]\sim\mathcal{CN}(\mathbf{0}, \sigma^2\mathbf{I}_{N_B})$ is the additive noise vector. Let $N_{\text{beam}} = N_{\text{slot}}N_{\text{RF}}$ denote the total number of pilot beams. After $N_{\text{slot}}$ training slots, the BS acquires the measurement vector $\bar{\mathbf{y}}[s] \triangleq [\mathbf{y}^T_1[s], \dots, \mathbf{y}^T_{N_{\text{slot}}}[s]]^T \in\mathbb{C}^{N_{\text{beam}} \times 1}$ for $\mathbf{h}[s]$ as
\begin{align}\label{eq:general_expression_training}
\bar{\mathbf{y}}[s]  &= 
\sqrt{P_p}\begin{bmatrix}
\mathbf{W}^H_1[s]\\
\vdots \\
\mathbf{W}^H_{N_{\text{slot}}}[s]
\end{bmatrix} \mathbf{h}[s] 
+
\begin{bmatrix}
\mathbf{W}^H_1[s]\mathbf{n}_1[s]\\
\vdots \\
\mathbf{W}^H_{N_{\text{slot}}}[s]\mathbf{n}_{N_{\text{slot}}}[s]
\end{bmatrix}\nonumber \\[0.2cm]
&= \sqrt{P_p}\ \overline{\mathbf{W}}^H[s] \mathbf{h}[s] + \bar{\mathbf{n}}[s],
\end{align}
where $\overline{\mathbf{W}}[s] \triangleq [\mathbf{W}_1[s], \dots, \mathbf{W}_{N_\text{slot}}[s]]\in\mathbb{C}^{N_B\times N_{\text{beam}}}$, and $\bar{\mathbf{n}}[s]\in\mathbb{C}^{N_{\text{beam}}\times 1}$ denotes the effective noise. More particularly, $\mathbf{R}_{\bar{\mathbf{n}}[s]} \triangleq \sigma^2\text{diag}\left(\mathbf{W}^H_1[s]\mathbf{W}_1[s],\dots, \mathbf{W}^H_{N_\text{slot}}[s]\mathbf{W}_{N_\text{slot}}[s]\right)$ is the covariance matrix of the effective noise, which is colored in general. Regarding the pilot combiners, due to the hybrid array architecture, $\overline{\mathbf{W}}[s] = \overline{\mathbf{W}}_{\text{RF}}\overline{\mathbf{W}}_{\text{BB}}[s]$, with $\overline{\mathbf{W}}_{\text{RF}}=[\mathbf{W}_{\text{RF},1}, \dots, \mathbf{W}_{\text{RF}, N_{\text{slot}}}]\in\mathbb{C}^{N_B\times N_{\text{beam}}}$ containing the RF pilot beams and $\overline{\mathbf{W}}_{\text{BB}}[s] = \text{blkdiag}(\mathbf{W}_{\text{BB},1}[s], \dots, \mathbf{W}_{\text{BB}, N_{\text{slot}}}[s])\in\mathbb{C}^{N_{\text{beam}}\times N_{\text{beam}}}$ comprising the $N_{\text{RF}}\times N_{\text{RF}}$ baseband combiners. The design of the pilot combiners is detailed in Section~\ref{sec_pilot_beamdesign}.

\subsection{Least Squares Estimator}
From~\eqref{eq:general_expression_training}, we have $N_{\text{beam}}$ observations, while $\mathbf{h}[s]$ includes $N_B$ variables. Thus, to obtain a good estimate of $\mathbf{h}[s]$, we need that $N_{\text{beam}} \geq N_B$. With this condition, the \ac{LS} estimate is\footnote{We consider the \ac{LS} instead of the minimum mean-square error (MMSE) method because we focus on estimators that exploit only instantaneous \ac{CSI}.}
\begin{equation}\label{eq:ls_estimator}
\hat{\mathbf{h}}^{\text{LS}}[s] = \mathbf{Q}^{\dagger}_s\bar{\mathbf{y}}[s],
\end{equation}
where $\mathbf{Q}_s \triangleq \sqrt{P_p} \ \overline{\mathbf{W}}^H[s]\in\mathbb{C}^{N_{\text{beam}}\times N_B}$ is the sensing matrix. The mean square error (MSE) of the LS estimator for the $s$th subcarrier is given by
\begin{align}
J_s^{\text{LS}} \triangleq \mathbb{E}\left\{ \left \|\mathbf{h}[s] - \hat{\mathbf{h}}^{\text{LS}}[s] \right \|^2 \right\} = \text{tr}\left(\mathbf{Q}_s^{\dagger}\mathbf{R}_{\bar{\mathbf{n}}[s]}(\mathbf{Q}^{\dagger}_s)^{H}\right).
\end{align}
The optimal $\mathbf{Q}_s$ satisfies $\mathbf{Q}^H_s\mathbf{Q}_s = P_p\mathbf{I}_{N_B}$~\cite{mmWave_ps_or_sw, chs_new_approach}. In the hybrid array architecture under consideration, this is achieved by $\overline{\mathbf{W}}_{\text{BB}}[s] = \mathbf{I}_{N_B}$ and $\overline{\mathbf{W}}_{\text{RF}} = \mathbf{U}\in\mathbb{C}^{N_B\times N_B}$, where $\mathbf{U}$ is the \ac{DFT} matrix generating the RF pilot beams~\cite{mmWave_ps_or_sw}. We then have $\mathbf{R}_{\bar{\mathbf{n}}[s]} =\sigma^2 \mathbf{I}_{N_B}$, $\mathbf{Q}_s^{\dagger} = (1/\sqrt{P_p})\mathbf{U}$, and
\begin{equation}\label{eq:mse_ls_fulltraining}
J_s^{\text{LS}} = \sigma^2 N_B/P_p.
\end{equation}
The LS estimator~\eqref{eq:ls_estimator} requires $N_{\text{beam}} \geq N_B$, and hence yields a prohibitively high training overhead when the number of \ac{RF} chains is much smaller than the number of \ac{BS} antennas.
\subsection{Sparse Formulation and Orthogonal Matching Pursuit}
By exploiting the angular sparsity of THz channels, we can have a sparse formulation of the channel estimation problem as follows. The physical channel in~\eqref{eq:physical_model} is also expressed as
\begin{equation}
\mathbf{h}[s] = \mathbf{A}[s]\bm{\beta}[s],
\end{equation}
where $\mathbf{A}[s] \triangleq [\mathbf{a}(\phi_0,\theta_0, f_s),\dots, \mathbf{a}(\phi_L,\theta_L, f_s) ]\in~\mathbb{C}^{N_B\times (L+1)}$, with $\mathbf{a}(\phi_l,\theta_l, f_s)$ being specified by \eqref{eq:upa_response} for $f=f_s$, is the so-called wideband array response matrix, and $\bm{\beta}[s] \triangleq [\beta_0(f_s)e^{-j2\pi f_s \tau_0}, \dots, \beta_L(f_s)e^{-j2\pi f_s \tau_L}]^T\in\mathbb{C}^{(L+1)\times 1}$ is the vector of channel gains. Next, consider a dictionary $\bar{\mathbf{A}}[s]\in~\mathbb{C}^{N_B\times G}$ whose $G$ columns are the array response vectors associated with a predefined set of \ac{DoA}. Then, the uplink channel can be approximated as
\begin{equation}
\mathbf{h}[s]\approx  \bar{\mathbf{A}}[s]\bar{\bm{\beta}}[s],
\end{equation}
where $\bar{\bm{\beta}}[s]\in\mathbb{C}^{G\times 1}$ has $L+1$ nonzero entries whose positions and values correspond to their \ac{DoA} and path gains~\cite{overview_sign_mmwave}. Therefore,~\eqref{eq:general_expression_training} is recast as
\begin{align}\label{eq:sparse_rx_pilot_signal}
\bar{\mathbf{y}}[s] = \mathbf{\Phi}_s \bar{\bm{\beta}}[s]  + \bar{\mathbf{n}}[s],
\end{align}
where $\mathbf{\Phi}_s\triangleq \sqrt{P_p}\ \overline{\mathbf{W}}^H[s]\bar{\mathbf{A}}[s] \in\mathbb{C}^{N_{\text{beam}}\times G}$ is the \textit{equivalent} sensing matrix. Since $(L+1)\ll G$, the channel gain vector $\bar{\bm{\beta}}[s]$ is $(L+1)$-sparse, and the channel estimation problem can be formulated as the sparse recovery problem~\cite{mmWave_ps_or_sw}
\begin{align}\label{eq:opt_problem}
\hat{\bar{\bm{\beta}}}[s] = \arg&\min_{\bar{\bm{\beta}}[s]} \  \|\bar{\bm{\beta}}[s] \|_1 \nonumber\\
&\text{s.t.} \quad \left \| \bar{\mathbf{y}}[s] -\mathbf{\Phi}_s \bar{\bm{\beta}}[s] \right \|_2 \leq \epsilon
\end{align}
where $\epsilon \leq \mathbb{E}\{\|\bar{\mathbf{n}}[s]\|_2\}$ is an appropriately chosen bound on the mean magnitude of the effective noise. The above optimization problem can be solved for each subcarrier independently, i.e., single measurement vector formulation. Finally, the estimate of $\mathbf{h}[s]$ is obtained as $\hat{\mathbf{h}}^{\text{CS}}[s] =  \bar{\mathbf{A}}[s]\hat{\bar{\bm{\beta}}}[s]$.

Several greedy algorithms have been proposed to find approximate solutions of the $l_1$-norm optimization problem. The orthogonal matching pursuit (OMP) algorithm~\cite{omp} described in Algorithm 1 is one of the most common and simple greedy CS methods that can solve~\eqref{eq:opt_problem}.
\begin{algorithm}[H]
	\caption{OMP-Based Estimator}
	
	\begin{algorithmic}[1]
		\Statex Input: equivalent sensing matrix $\mathbf{\Phi}_s$ and measurement vector $\bar{\mathbf{y}}[s]$ for the $s$th subcarrier, and a threshold $\epsilon$. 
		\State $\mathcal{I}_{-1} = \emptyset$, $\mathcal{G} = \{1,\dots, G\}$, $\mathbf{r}_{-2}[s] = \mathbf{0}$, $\mathbf{r}_{-1}[s] = \bar{\mathbf{y}}[s]$, and $l=0$.
		\While{$\|\mathbf{r}_{l-1}[s] - \mathbf{r}_{l-2}[s]\|^2 > \epsilon$} 
		\State $g^\star = \underset{g\in\mathcal{G}}{\arg\max} \  \left|\mathbf{\Phi}_s^H(g)\mathbf{r}_{l-1}[s]\right |$
		\State $\mathcal{I}_{l} = \mathcal{I}_{l-1}\cup\{g^{\star}\}$ 
		\State $\mathbf{r}_l[s] = \left(\mathbf{I}_{N_{\text{beam}}} -  \mathbf{\Phi}_s(\mathcal{I}_l) \mathbf{\Phi}_s^{\dagger}(\mathcal{I}_l)\right)\bar{\mathbf{y}}[s]$
		\State $l = l+1$ 
		\EndWhile 
		\State $\hat{\bar{\bm{\beta}}}[s] = \mathbf{\Phi}_s^{\dagger}(\mathcal{I}_{l-1})\bar{\mathbf{y}}[s]$
		\State \textbf{return} $\hat{\mathbf{h}}^{\text{CS}}[s]=  \bar{\mathbf{A}}[s]\hat{\bar{\bm{\beta}}}[s]$.
	\end{algorithmic}\label{algo_omp}
\end{algorithm}
 \begin{figure*}[t]
	\centering	
	\includegraphics[width=0.65\linewidth]{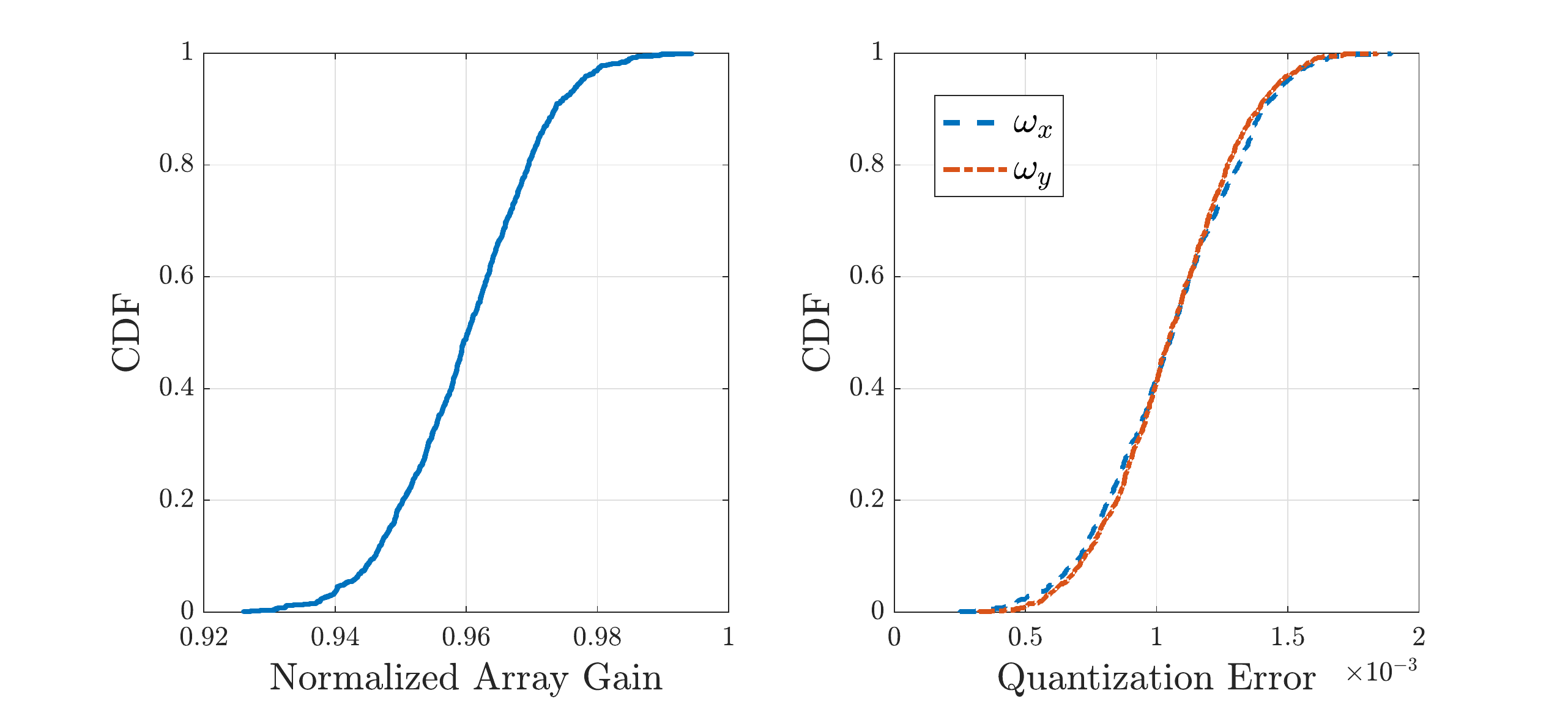}
	\caption{\ac{CDF} of the normalized array gain and quantization error for a single-path channel and a super-resolution dictionary with $G_x =4 N$ and $G_y = 4M$; 1,000 channel realizations, $40\times 40$-element \ac{UPA}, $\mathbf{f}_{\text{RF}} = (1/\sqrt{N_B})\mathbf{a}(\bar{\omega}_x(q),\bar{\omega}_y(p), f_s)$, $B = 40$ GHz, $S=400$ subcarriers, and $s=200$th subcarrier.}
	\label{fig:Fig_cdf}
\end{figure*} 

\subsection{Proposed Channel Estimator}\label{sec:proposed_ce}
\subsubsection{Wideband Dictionary for UPAs} For half-wavelength antenna separation, the array response vector~\eqref{eq:upa_response} is recast as 
\begin{multline}\label{eq:upa_response_sf}
\mathbf{a}(\omega_x,\omega_y,f) = \left[ 1, \dots,e^{-j2\pi  \left(1+\frac{f}{f_c}\right)(n \omega_x+ m\omega_y)} \right ., \\
\dots, \left . e^{-j2\pi \left(1+\frac{f}{f_c}\right)\left((N-1) \omega_x+ (M-1)\omega_y\right)}\right]^T,
\end{multline}
where $\omega_x = 1/2\sin\theta\cos\phi$ and $\omega_y = 1/2\sin\theta\sin\phi$ are the \textit{spatial frequencies}~\cite{opt_array_processing}. The one-to-one mapping between the spatial frequencies $(\omega_x, \omega_y)$ and the physical angles $(\phi, \theta)$ is given by the relationships
\begin{align}\label{eq:spatial_fre_physical_angles1}
\phi &= \text{tan}^{-1}\left(\omega_y/\omega_x\right), \\ \label{eq:spatial_fre_physical_angles2}
\theta &= \text{sin}^{-1}\left(2\sqrt{\omega_x^2 + \omega_y^2}\right).
\end{align}
Since both $\omega_x$ and $\omega_y$ lie in $[-1/2,1/2]$, we can consider the grids of discrete spatial frequencies
\begin{align}
\mathcal{G}_x &= \left\{\bar{\omega}_x(q) = q/G_x, \ q=-(G_x-1)/2,\dots, (G_x-1)/2\right\}, \\
\mathcal{G}_y&=\{\bar{\omega}_y(p) = p/G_y, \ p=-(G_y-1)/2,\dots, (G_y-1)/2\},
\end{align}	
where $G_xG_y = G$ is the overall dictionary size. 

For the $s$th subcarrier, we define the array response matrices $\bar{\mathbf{A}}_x[s]\in\mathbb{C}^{N\times G_x}$ and $\bar{\mathbf{A}}_y[s]\in\mathbb{C}^{M\times G_y}$ whose columns are the array response vectors $\mathbf{a}_x(\cdot, f_s)$ and $\mathbf
a_y(\cdot,f_s)$ evaluated at the grid points of $\mathcal{G}_x$ and $\mathcal{G}_y$, respectively. Now, the dictionary $\bar{\mathbf{A}}[s]\triangleq\bar{\mathbf{A}}_x[s]\otimes\bar{\mathbf{A}}_y[s]\in\mathbb{C}^{N_B\times G}$ can be used to approximate the uplink channel $\mathbf{h}[s]$ at the $s$th subcarrier. Although this approximation entails quantization errors, they become small for large $G_x$ and $G_y$~\cite{overview_sign_mmwave}. More specifically, we can use a super-resolution dictionary with $G_x > N$ and $G_y > M$ to reduce the mismatch between the quantized and the actual channel. We evaluate the accuracy of the proposed dictionary by generating a \ac{DoA} with $(\omega_x,\omega_y)$, which is then quantized to the closest value $(\bar{\omega}_x(q),\bar{\omega}_y(p))$. Figure~\ref{fig:Fig_cdf} shows the \acf{CDF} of the normalized array gain $\left|\mathbf{a}^H(\bar{\omega}_x(q),\bar{\omega}_y(p), f_s)\mathbf{a}(\omega_x,\omega_y, f_s)\right|^2/N_B^2$, and the quantization errors $|\omega_x - \bar{\omega}_x(q)|$ and $|\omega_y - \bar{\omega}_y(p)|$ of the spatial frequencies. As we observe, the errors are small, and do not affect significantly the normalized array gain. Consequently, we can neglect the quantization errors, and assume that the \ac{DoA} of each path lies on the dictionary grid. Note that for $G_x = N$  and $G_y=M$, the dictionary $\bar{\mathbf{A}}[s]$ reduces to the known virtual channel representation (VCR)~\cite{VCR} in the spatial-narrowband case. Lastly, a similar representation, termed extended VCR, was introduced in~\cite{unified_ttd/fdd} for narrowband massive MIMO systems. 

\subsubsection{Generalized Multiple Measurement Vector Problem} Due to the frequency-dependent dictionary, the channel gain vectors $\{\bar{\bm{\beta}}[s]\}_{s=0}^{S-1}$ share the same support. Therefore, we can exploit the common support property and consider the problem in~\eqref{eq:opt_problem} as a generalized multiple measurement vector (GMMV) problem, where multiple sensing matrices are employed~\cite{ce_fdd_massive_mimo}. To tackle the GMMV problem, we employ the simultaneous OMP algorithm~\cite{somp}. The proposed channel estimator is described in Algorithm~2.
\begin{algorithm}[H]
	\caption{GSOMP-Based Estimator}
	
	\begin{algorithmic}[1]
		\Statex Input: set $\mathcal{S}$ of pilot subcarriers, sensing matrices $\mathbf{\Phi}_s$ and measurement vectors $\bar{\mathbf{y}}[s], \forall s\in\mathcal{S}$, and a threshold $\epsilon$. 
		\State $\mathcal{I}_{-1} = \emptyset$, $\mathcal{G} = \{1,\dots, G\}$, $\mathbf{r}_{-1}[s] = \bar{\mathbf{y}}[s]$, $\mathsf{MSE} = \sum_{s\in\mathcal{S}}\|\bar{\mathbf{y}}[s]\|^2$, and $l=0$.
		\While{$\mathsf{MSE} > \epsilon$}
		\State $g^\star = \underset{g\in\mathcal{G}\setminus \mathcal{I}_{l-1}}{\arg\max} \  \underset{s\in\mathcal{S}}{\sum} \left|\mathbf{\Phi}_s^H(g)\mathbf{r}_{l-1}[s]\right |$
		\State $\mathcal{I}_{l} = \mathcal{I}_{l-1}\cup\{g^{\star}\}$ 
		\State $\mathbf{r}_l[s] = \left(\mathbf{I}_{N_{\text{beam}}} -  \mathbf{\Phi}_s(\mathcal{I}_l) \mathbf{\Phi}_s^{\dagger}(\mathcal{I}_l)\right)\bar{\mathbf{y}}[s], \ \forall s\in\mathcal{S}$
		\State  $\mathsf{MSE} = \frac{1}{|\mathcal{S}|}\sum_{s\in\mathcal{S}}\| \mathbf{r}_l[s] - \mathbf{r}_{l-1}[s]\|^2$
		\State $l = l+1$ 
		\EndWhile
		\State $\hat{\bar{\bm{\beta}}}[s] = \mathbf{\Phi}_s^{\dagger}(\mathcal{I}_{l-1})\bar{\mathbf{y}}[s], \forall s\in\mathcal{S}$
		\State \textbf{return} $\hat{\mathbf{h}}^{\text{CS}}[s]=  \bar{\mathbf{A}}[s]\hat{\bar{\bm{\beta}}}[s], \forall s\in\mathcal{S}$.
	\end{algorithmic}\label{algo_gsomp}
\end{algorithm} 
Regarding the stopping criterion of the OMP/GSOMP algorithm, we design the pilot combiners so that the effective noise is white. In this case, the variance of the noise power is $\mathbb{E}\left\{\|\bar{\mathbf{n}}[s]\|^2\right\} = N_{\text{beam}}\sigma^2$, and the threshold can be chosen as $\epsilon = N_{\text{beam}}\sigma^2$, or a fraction of the average noise power. Additionally, a thresholding step can be incorporated into the algorithms, in which only the entries of the estimate $\hat{\bar{\bm{\beta}}}$ with power higher than the noise variance will be selected as detected paths. After estimating the spatial frequencies of each path, the physical angles are obtained through~\eqref{eq:spatial_fre_physical_angles1} and \eqref{eq:spatial_fre_physical_angles2}, which are then used in the TTD-based wideband combiner.

\subsubsection{Pilot Beam Design}\label{sec_pilot_beamdesign}
The elements of the RF combiner $\overline{\mathbf{W}}_{\text{RF}}$ are selected from the set $\{-1/\sqrt{N_B}, 1/\sqrt{N_B}\}$ with equal probability. The reason we adopt a randomly formed RF combiner is that it has been shown to have a low mutual-column coherence, and therefore can be expected to attain a high recovery probability according to the compressive sensing theory~\cite{omp_rand_measurements}. The specific RF pilot design leads to a colored effective noise, however the SOMP algorithm is based on the assumption that the noise covariance matrix is diagonal. To this end, we design the baseband combiner such that the combined noise remains white. In particular, let $\mathbf{D}^H_t\mathbf{D}_t$ be the Cholesky decomposition of $\mathbf{W}^H_{\text{RF,t}}\mathbf{W}_{\text{RF,t}}$, where $\mathbf{D}\in\mathbb{C}^{N_{\text{RF}}\times N_{\text{RF}}}$ is an upper triangular matrix. Then, the baseband combiner of the $t$th slot is set to $\mathbf{W}_{\text{BB},t}[s] = \mathbf{D}^{-1}_t$, and hence $\overline{\mathbf{W}}[s] = \overline{\mathbf{W}}_{\text{RF}}\text{blkdiag}(\mathbf{D}^{-1}_1,\dots, \mathbf{D}^{-1}_{N_{\text{slot}}})$. Under this pilot beam design, the covariance matrix of the effective noise becomes $\mathbf{R}_{\bar{\mathbf{n}}} = \sigma^2\mathbf{I}_{N_{\text{beam}}}$, yielding the desired result. We finally point out that the combiners $\overline{\mathbf{W}}[s]$ can be computed offline. 

\subsection{Performance of the Proposed Channel Estimator}
\subsubsection{Lower Bound Error Analysis} As previously mentioned, for semi-unitary combiners $\mathbf{W}_t[s]$ with $\mathbf{W}_t^H[s]\mathbf{W}_t[s] = \mathbf{I}_{N_\text{RF}}, \forall t=1,\dots, N_{\text{slot}}$, the covariance matrix of the effective noise $\bar{\mathbf{n}}[s]$ is equal to $\sigma^2\mathbf{I}_{N_{\text{beam}}}$. Next, we derive the \ac{CRLB} assuming that the \ac{GSOMP} recovers the exact support of $\bar{\bm{\beta}}[s]$, i.e., $\mathcal{I}_{l-1} = \text{supp}\left(\bar{\bm{\beta}}[s]\right)$ = $\mathcal{I}$.\footnote{This is a well accepted assumption in the related literature; see~\cite{ce_frequency_selective_mmWave} and references therein.} To this end, we can define the following linear model for the $s$th subcarrier
\begin{equation}\label{eq:glm_model}
\bar{\mathbf{y}}[s] = \mathbf{\Phi}_s(\mathcal{I})\tilde{\bar{\bm{\beta}}}[s] + \bar{\mathbf{n}}[s],
\end{equation}
where $\tilde{\bar{\bm{\beta}}}[s]\in\mathbb{C}^{L\times 1}$ denotes the vector to be estimated, and $\bar{\mathbf{y}}[s]$ is distributed as $\mathcal{CN}\left(\mathbf{\Phi}_s(\mathcal{I})\tilde{\bar{\bm{\beta}}}[s], \sigma^2\mathbf{I}_{N_{\text{beam}}}\right)$. The model in~\eqref{eq:glm_model} is linear on the parameter vector $\tilde{\bar{\bm{\beta}}}[s]$, and the solution $\hat{\bar{\bm{\beta}}}[s] = \mathbf{\Phi}_s^{\dagger}(\mathcal{I})\bar{\mathbf{y}}[s]$ gives $\mathbb{E}\left\{\hat{\bar{\bm{\beta}}}[s]\right\} = \tilde{\bar{\bm{\beta}}}[s]$. Specifically, $\hat{\bar{\bm{\beta}}}[s]$ is the mininum variance unbiased estimator of $\tilde{\bar{\bm{\beta}}}[s]$, hence attaining the CRLB~\cite{fund_statistical_processing}. Next, the Fisher information matrix for~\eqref{eq:glm_model} is calculated as
\begin{equation}
\mathbf{I}\left(\tilde{\bar{\bm{\beta}}}[s]\right) = \frac{1}{\sigma^2} \mathbf{\Phi}^H_s(\mathcal{I}) \mathbf{\Phi}_s(\mathcal{I}).
\end{equation}
The channel estimate for the $s$th subcarrier is acquired as $\hat{\mathbf{h}}^{\text{CS}}[s] = \bar{\mathbf{A}}_s(\mathcal{I})\hat{\bar{\bm{\beta}}}[s]$, where $\bar{\mathbf{A}}_s(\mathcal{I})$ denotes the matrix with the columns of $\bar{\mathbf{A}}[s]$ given by the support $\mathcal{I}$. Let $J^{\text{CS}}_s$ denote the MSE of the OMP. Since $\mathbb{E}\left\{\hat{\mathbf{h}}^{\text{CS}}\right\} = \bar{\mathbf{A}}_s(\mathcal{I})\tilde{\bar{\bm{\beta}}}[s] \triangleq \bm{\psi}\left(\tilde{\bar{\bm{\beta}}}[s]\right)$, the CRLB for the $s$th subcarrier is given by~\cite{fund_statistical_processing}
\begin{align}\label{eq:crlb}
J^{\text{CS}}_s  \geq \text{tr}\left\{\ \frac{\partial\bm{\psi}\left(\tilde{\bar{\bm{\beta}}}[s]\right)}{\partial\tilde{\bar{\bm{\beta}}}[s]}\mathbf{I}^{-1}\left(\tilde{\bar{\bm{\beta}}}[s]\right) \frac{\partial\bm{\psi}^H\left(\tilde{\bar{\bm{\beta}}}[s]\right)}{\partial\tilde{\bar{\bm{\beta}}}[s]}\right\},
\end{align}
where $\partial\bm{\psi}\left(\tilde{\bar{\bm{\beta}}}[s]\right)/\partial\tilde{\bar{\bm{\beta}}}[s] =\bar{\mathbf{A}}_s(\mathcal{I})$.

\subsubsection{Complexity Analysis} In this section, we  detail the computational complexity per iteration $l$ of the GSOMP scheme. Specifically, we have the following operations: 
\begin{itemize}
\item The $l_2$-norm operations at step $1$ and step $6$ have $\mathcal{O}(|\mathcal{S}|N_{\text{beam}})$ complexity.
\item The calculation of the product $\mathbf{\Phi}_s^H(g)\mathbf{r}_{l-1}[s]$ at step $3$ is $\mathcal{O}(|\mathcal{S}|N_{\text{beam}}(G-l))$ because there are $G-l$ elements to examine at the $l$th iteration, where $G$ is the size of the dictionary.
\item To find the maximum element from $G-l$ values at step~$3$ is on the order of $\mathcal{O}(G-l)$.
\item The \ac{LS} operation at step $5$ is $\mathcal{O}(l^3 + 2l^2N_{\text{beam}})$ for each pilot subcarrier. This is because $\mathbf{\Phi}(\mathcal{I}_l)$ is a $N_{\text{beam}}\times l$ matrix, and hence its pseudoinverse entails $l^3 + l^2N_{\text{beam}}$ operations plus the multiplication with $\mathbf{\Phi}(\mathcal{I}_l)$ entailing $l^2N_{\text{beam}}$ additional multiplications. 
\end{itemize}
Given the above, the overall online computational complexity is $\mathcal{O}\left(|\mathcal{S}|(N_{\text{beam}}(G-l) + l^3 + 2l^2N_{\text{beam}}) + (G-l)\right)$. Note that the \ac{OMP} has $\mathcal{O}(|\mathcal{S}|G)$ at step $3$ for finding the maximum correlation between the measurement vector and the columns of the dictionary. As a result, the GSOMP leads to a computational reduction as well. 

\section{The Multi-Antenna User Case}\label{sec:multiantenna_users}
We now discuss how the previous analysis can be extended to the case of a multi-antenna user. To this end, we consider a user with an $N_U$-element ULA. The frequency response of the uplink channel, $\mathbf{H}(f)\in\mathbb{C}^{N_B\times N_U}$, is then expressed as
\begin{equation}
\mathbf{H}(f) = \sum_{l=0}^{L}\beta_l(f) \mathbf{a}_B(\phi_l,\theta_l, f)\mathbf{a}^H_U(\varphi_l, f)e^{-j2\pi f\tau_l},
\end{equation}
where $\mathbf{a}_B(\cdot,\cdot,\cdot)$ denotes the response vector \eqref{eq:upa_response} of the \ac{BS} array, $\varphi_l$ is the \ac{AoD} of the $l$th path from the user, and
\begin{multline}
\mathbf{a}_U(\varphi, f) \triangleq \left[1, e^{-j2\pi \left(f_c + f\right)\frac{d}{c}\sin\varphi}, \right . \\
\left .\dots, e^{-j2\pi (f_c +f)(N_U-1)\frac{d}{c}\sin\varphi}\right]^T
\end{multline}
is the wideband response vector of the user array. 

At the \ac{BS}, the post-processed baseband signal for the $s$th subcarrier is expressed as 
\begin{equation}
	\mathbf{y}[s] =\mathbf{F}^H[s]\left(\mathbf{H}[s]\mathbf{B}[s]\tilde{\mathbf{x}}[s] + \mathbf{n}[s]\right),
\end{equation}
where $\mathbf{B}[s] \in\mathbb{C}^{N_U\times N^u_{\text{RF}}}$ is the hybrid precoder when the user employs $N^u_{\text{RF}}$ \ac{RF} chains, $\tilde{\mathbf{x}}[s] = \mathbf{P}[s]\mathbf{x}[s]$ is the transmitted signal at the $s$th subcarrier, $\mathbf{P}[s] = \text{diag}(p_{1,s}, \dots, p_{N^u_{\text{RF}},s})$ is the power allocation matrix, and $\mathbf{x}[s]\sim\mathcal{CN}(\mathbf{0},\mathbf{I}_{N^u_{\text{RF}}})$ is the vector of data symbols. Furthermore, the power constraint $\sum_{s=0}^{S-1}\mathbb{E}\left\{\|\mathbf{B}[s]\tilde{\mathbf{x}}[s]\|^2\right\} \leq P_t$ should be satisfied, so that the transmit power does not exceed the user's power budget $P_t$.

\subsection{Hybrid Combining and Beamforming}\label{sec:hybrid_mu}
Consider a single-path channel with \ac{AoD} $\varphi$ from the user and \ac{DoA} $(\phi,\theta)$ at the \ac{BS}. For the frequency-flat beamformer $(1/\sqrt{N_U})\mathbf{a}_U(\varphi, 0)$ and combiner $(1/\sqrt{N_B})\mathbf{a}_B(\phi, \theta, 0)$, the normalized array gain in~\eqref{eq:array_gain} is recast as in~\eqref{eq:normalized_ag_mu} at the bottom of this page, where $\Delta(\varphi)\triangleq d \sin\varphi/ c$. Employing TTD-based combining and beamforming yields $G(\phi,\theta,\varphi,f)\approx~1$, and the \ac{SNR} at the $s$th subcarrier is approximately equal to $|\beta(f_s)|^2 N_U N_BP_d/\sigma^2$. Compared to the single-antenna user case, we have an additional beamforming gain $N_U$. 
\begin{figure*}[b]
		\hrulefill
		\begin{align}\label{eq:normalized_ag_mu}
		G(\phi,\theta, \varphi, f) &=\frac{|\mathbf{a}_B^H(\phi,\theta,0)\mathbf{a}_B(\phi,\theta,f)|^2}{N_B^2}\frac{|\mathbf{a}_U^H(\varphi, f)\mathbf{a}_U(\varphi, 0)|^2}{N_U^2}  \nonumber \\
		&= \left|D_{N}(2\pi f\Delta_x(\phi,\theta))\right|^2  \left | D_{M}(2\pi f\Delta_y(\phi,\theta)) \right|^2\left|D_{N_U}(2\pi f\Delta(\varphi))\right|^2.
		\end{align}
\end{figure*}

Now consider, for  instance, a multi-path channel of $L=2$ \ac{NLoS} paths. In a fully-digital array, the combiner and precoder maximizing the achievable rate are given by the \ac{SVD} of the channel matrix $\mathbf{H}[s]$~\cite{mmWave_bsq_ref2}. For our hybrid analog-digital array structure, we adopt a practical approach, as in~\cite{Hybrid_TD_precoding}. We first decompose the channel matrix as $\mathbf{H}(f) = \mathbf{H}_B(f)\mathbf{H}^H_{U}(f)$, where 
\begin{equation}\label{eq:H_b}
\mathbf{H}_{B}(f) =
\begin{bmatrix}
\mathbf{a}_B(\phi_1,\theta_1, f), & \mathbf{a}_B(\phi_2,\theta_2, f)
\end{bmatrix},
\end{equation}
and 
\begin{multline}
\mathbf{H}_U(f) = \left[\beta_1(f)\mathbf{a}_{U}(\varphi_1, f)e^{-j2\pi f\tau_1}, \right . \\ \left .\beta_2(f)\mathbf{a}_{U}(\varphi_2, f)e^{-j2\pi f\tau_2}\right].
\end{multline}
Next, the \ac{RF} combiner and beamformer are the matched filters of the channels $\mathbf{H}_B(f)$ and $\mathbf{H}^H_U(f)$, respectively, whereas the baseband combiner and precoder are designed using the \ac{SVD} of the effective channel, when both ends have full~\ac{CSI}. Note that for a multi-path channel with $L > N_{\text{RF}}$ paths, the user communicates at most $\min(L, N_{\text{RF}})$ spatial streams to the \ac{BS} in the absence of inter-stream interference through \ac{SVD}-based transmission.

\subsection{Sparse Channel Estimation}
The user employs a training codebook $\{\mathbf{v}_i\in\mathbb{C}^{N_U\times 1}, i=1,\dots, N^u_{\text{beam}}\}$, which consists of $N_{\text{beam}}^u$ pilot \ac{RF} beamformers. When the $i$th pilot beamformer is used during $N_{\text{slot}}$ training slots,~\eqref{eq:general_expression_training} is recast as 
\begin{equation}
\bar{\mathbf{y}}_i[s] =  \sqrt{P_p}\ \overline{\mathbf{W}}^H[s] \mathbf{H}[s]\mathbf{v}_i + \bar{\mathbf{n}}_i[s].
\end{equation}
By collecting all vectors $\bar{\mathbf{y}}_i[s] $ into a single matrix $\mathbf{Y}[s] = [\bar{\mathbf{y}}_1[s],\dots,\bar{\mathbf{y}}_{N^u_{\text{beam}}}[s]]\in\mathbb{C}^{N_{\text{beam}} \times N^u_{\text{beam}}}$, we can write  
\begin{equation}\label{eq:training_multiantenna_user}
\mathbf{Y}[s] =  \sqrt{P_p}\ \overline{\mathbf{W}}^H[s] \mathbf{H}[s]\mathbf{V} + \mathbf{N}[s],
\end{equation}
where $\mathbf{V} = [\mathbf{v}_1,\dots, \mathbf{v}_{N^u_{\text{beam}}} ]\in\mathbb{C}^{N_U\times N^u_{\text{beam}}}$, and $ \mathbf{N} = [\bar{\mathbf{n}}_1[s],\dots,\bar{\mathbf{n}}_{N^u_{\text{beam}}}[s]]\in\mathbb{C}^{N_{\text{beam}} \times N^u_{\text{beam}}}$. Utilizing the identity $\text{vec}(\mathbf{A}\mathbf{B}\mathbf{C}) =(\mathbf{C}^T\otimes \mathbf{A})\text{vec}(\mathbf{B})$, we express~\eqref{eq:training_multiantenna_user} in vector form as
\begin{equation}
\text{vec}(\mathbf{Y}[s]) = \sqrt{P_p}\left(\mathbf{V}^T\otimes \overline{\mathbf{W}}^H[s]\right)\text{vec}(\mathbf{H}[s]) + \text{vec}(\mathbf{N}[s]),
\end{equation}
where $\text{vec}(\mathbf{Y}[s])\in\mathbb{C}^{N_{\text{beam}}N^u_{\text{beam}}\times 1}$ is the overall measurement vector, $\text{vec}(\mathbf{H}[s])\in\mathbb{C}^{N_BN_U\times 1}$ is the uplink channel to be estimated, and $\text{vec}(\mathbf{N}[s])\in\mathbb{C}^{N_{\text{beam}}N^u_{\text{beam}}\times 1}$ is the noise vector. Now, the proposed GSOMP-based estimator can readily be used by considering the equivalent sensing matrix $\mathbf{\Phi}_s = \sqrt{P_p}\left(\mathbf{V}^T\otimes \overline{\mathbf{W}}^H[s]\right)\bar{\mathbf{A}}[s] \in\mathbb{C}^{N_{\text{beam}}N^u_{\text{beam}}\times GG^u}$, where $\bar{\mathbf{A}}[s] \triangleq \bar{\mathbf{A}}^*_u[s]\otimes(\bar{\mathbf{A}}_x[s]\otimes\bar{\mathbf{A}}_y[s])\in\mathbb{C}^{N_BN_U\times GG^u}$ is the equivalent dictionary accounting also for the dictionary $\bar{\mathbf{A}}_u[s]\in\mathbb{C}^{N_U\times G^u}$ of size $G^u$ at the user side. Finally, the estimated channel is constructed as $\text{vec}(\hat{\mathbf{H}}[s]) = \bar{\mathbf{A}}[s]\hat{\bar{\bm{\beta}}}[s]$.

\begin{table}[H]
	\centering
	\caption{Main Simulation Parameters~\cite{analytical_perf_thz, multiray_modeling_thz}}
	\label{Table:SimParameters}
	\begin{tabular}{|l  l|}
		\hline
		\textbf{Parameter} & \textbf{Value} \\
		\hline
		Bandwidth & $B=40$ GHz\\
		Carrier frequency & $f_c = 300$ GHz \\
		\hline
		Transmit power & $P_t=10$ dBm \\
		 Power density of noise& $\sigma^2 = -174$ dBm/Hz\\
		\hline
		Azimuth AoA  & 	 $\phi_l\sim\mathcal{U}\left(-\pi, \pi\right)$ \\
		Polar AoA & $\theta_l\sim\mathcal{U}\left(-\pi/2, \pi/2\right)$ \\
		\ac{LoS} path length & $\mathtt{D} = 15$ m \\
		\ac{ToA} of \ac{LoS} & $\tau_0=  50$ nsec \\
		\ac{ToA} of \ac{NLoS} & $\tau_{l}\sim\mathcal{U}(50, 55)$ nsec \\
		Absorption coefficient & $k_{\text{abs}}=0.0033 \ \text{m}^{-1}$\\
		Refractive index & $n_t = 2.24 - j0.025$\\
		Roughness factor & $\sigma_{\text{rough}} = 0.088\cdot 10^{-3}$ m\\
		\hline
	\end{tabular}
\end{table}
\begin{figure*}
	\centering
	\begin{subfigure}{.5\textwidth}
		\centering
		\includegraphics[width=.91\linewidth]{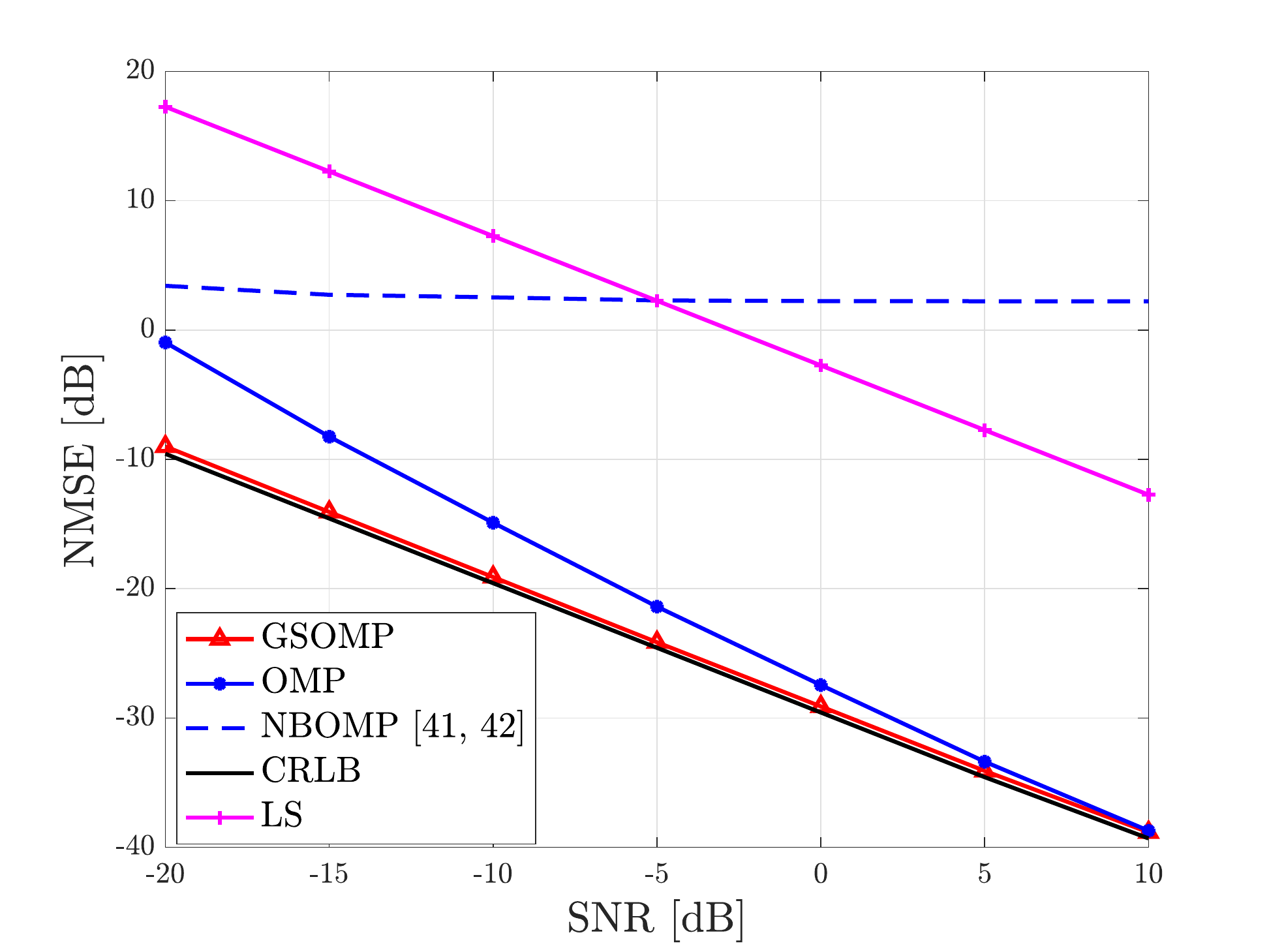}
		\caption{}
		\label{fig:NMSEvsSNR_a}
	\end{subfigure}%
	\begin{subfigure}{.5\textwidth}
		\centering
		\includegraphics[width=.91\linewidth]{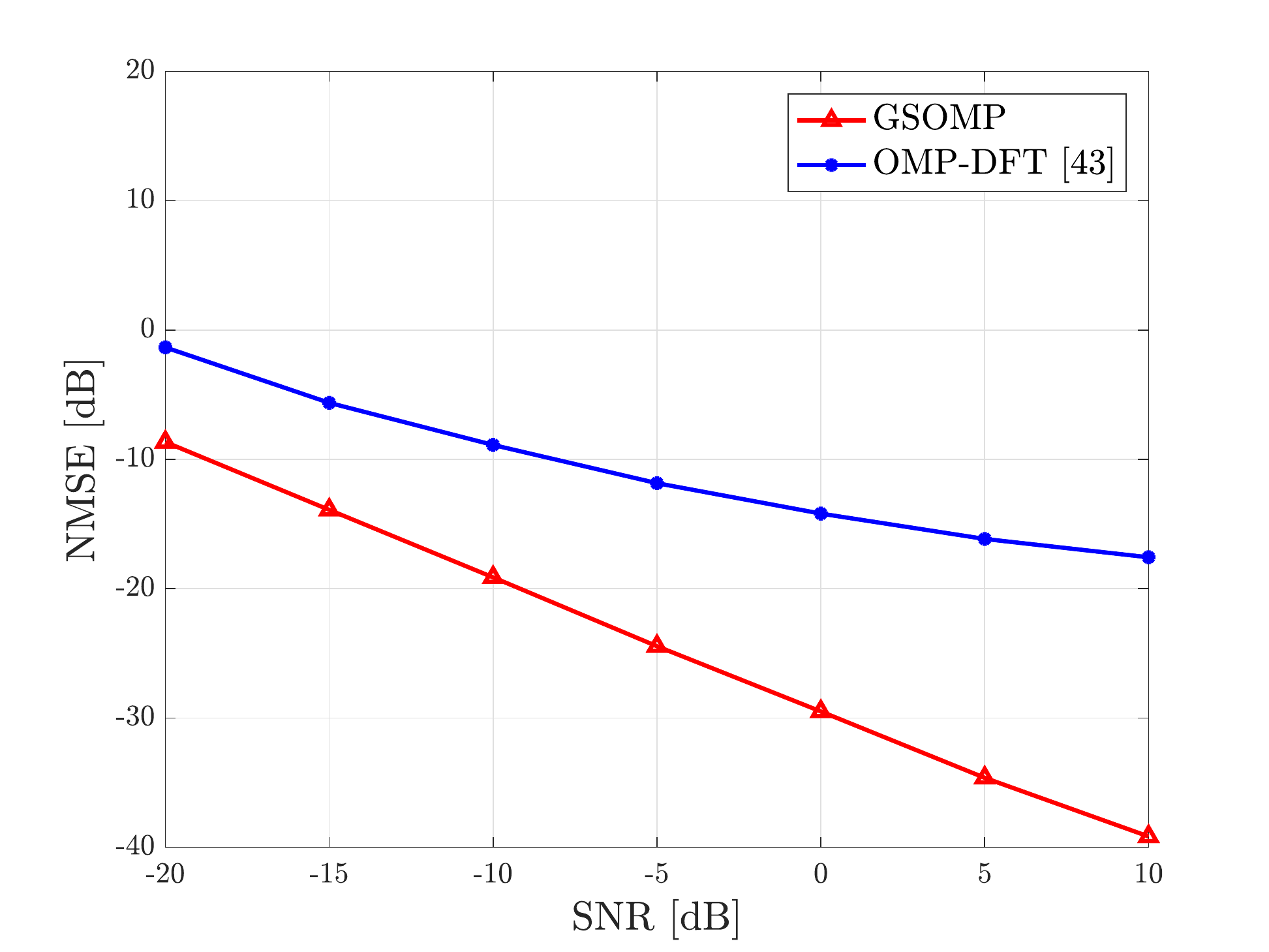}
		\caption{}
		\label{fig:NMSEvsSNR_b}
	\end{subfigure}
	\caption{NMSE versus SNR for a single-antenna user. The \ac{OMP}, NBOMP, and \ac{GSOMP} estimators are evaluated under partial training of $N_{\text{beam}} = 0.8N_B$ pilot beams; $40 \times 40$-element UPA, $N_{\text{RF}}=2$, \ac{NLoS} channel with $L=3$ paths, $S = 400$ subcarriers, and super-resolution dictionary with $G = 4N_B$.}
	\label{fig:NMSEvsSNR}
\end{figure*}

\section{Numerical Results}\label{sec:numerical_results}
In this section, we conduct numerical simulations to evaluate the performance of the proposed channel estimator and hybrid combiner. To this end, we consider the following setup:
\begin{itemize}
\item Number of OFDM Subcarriers: For a \ac{NLoS} multi-path scenario where $\tau_{l}\sim~\mathcal{U}(50, 55)$ nsec, the delay spread is $D_s = 5$ nsec. The coherence bandwidth is then calculated as $B_c = 1/(2D_s) = 100$~MHz~\cite{tse_book}, which results in $S\approx B/B_c = 400$ subcarriers. On the other hand, for a \ac{LoS} scenario, the delay spread is equal to the maximum delay across the \ac{UPA} due to the spatial-wideband effect. This results in $S \approx~18$ subcarriers for an $100\times 100$-element \ac{UPA} and $B = 40$~GHz.
\item Antenna Gain: Each \ac{BS} antenna element has a directional power pattern, $\Lambda(\phi,\theta)$, which is specified according to the 3GPP standard as~\cite{3GPP_tutorial}
\begin{equation}\label{eq:element_gain}
\Lambda(\phi,\theta) = \Lambda_{\max} - \min  \left[- \Lambda_H(\phi) - \Lambda_V(\theta), \Lambda_{\text{FBR}}\right],
\end{equation}
where 
\begin{align}\label{eq:gain_h}
\Lambda_H(\phi) &= - \min \left[ 12\left(\frac{\phi}{\phi_{\text{3dB}}}\right)^2, \Lambda_{\text{FBR}}\right], \\
\Lambda_V(\theta) &= - \min \left[ 12\left(\frac{\theta - 90\degree}{\theta_{\text{3dB}}}\right)^2, \text{SLA}_v\right],
\label{eq:gain_v}
\end{align}
where $\min\left[\cdot,\cdot\right]$ denotes the minimum between the input arguments, $\Lambda_{\max}$ is the maximum gain in the boresight direction, $\phi_{\text{3dB}} = 65\degree$ and $\theta_{\text{3dB}} = 65\degree$ are the horizontal and vertical half-power beamwidths, respectively, $\Lambda_{\text{FBR}} = 30$~dB is the front-to-back ratio, and $\text{SLA}_v = 30$ dB is the side lobe attenuation in the vertical direction. We choose $\Lambda_{\max} = 50$ dBi~\cite{analytical_perf_thz}. At the user side, we assume omnidirectional antennas. The channel model is then recast by replacing $\mathbf{a}(\phi,\theta,f)$ with $\sqrt{\Lambda(\phi,\theta)}\mathbf{a}(\phi,\theta,f)$~\cite{sparse_precoding}.
\end{itemize}
The other simulation parameters are summarized in Table~\ref{Table:SimParameters}. 

\subsection{Channel Estimation}

\subsubsection{Single-Antenna User}
Our main performance metric is the \ac{NMSE} versus the average receive SNR for the estimators intoduced previously. Specifically, for a given channel realization, the NMSE metric is defined as
\begin{equation}\label{eq:nmse}
\text{NMSE} \triangleq \frac{1}{|\mathcal{S}|} \sum_{s\in\mathcal{S}} \mathbb{E}\left\{\left\|\mathbf{h}[s] - \hat{\mathbf{h}}[s] \right\|^2 \big/ \|\mathbf{h}[s]\|^2\right\},
\end{equation}
where $\hat{\mathbf{h}}[s]$ denotes the estimate of the corresponding estimator. The NMSE is computed numerically over 100 channel realizations. The channel gains $\{\beta_l(f_s)\}_{l=1}^L$ are generated as $\mathcal{CN}(0, \sigma^2_{\beta})$, with $\sigma^2_{\beta} = 10^{-9}$, i.e., $-90$ dB, modeling the high path attenuation at THz frequencies~\cite{THz_Prop_models}.\footnote{The path gains are generated in this way in order to have a single average \ac{SNR} metric.} The average SNR is then calculated as $\text{SNR} = \sigma^2_{\beta}P_p/P_n$, where $P_p = P_t/|\mathcal{S}|$ is the power per pilot subcarrier, and $P_n = \Delta B \sigma^2$ is the noise power at each subcarrier, with $\Delta B \approx B/S$ being the subcarrier spacing. 

In the first numerical experiment, we compare the following estimation schemes:
\begin{itemize}
	\item The LS scheme under full training, i.e., $N_{\text{beam}} = N_B$.
	\item The narrowband OMP-based estimator (NBOMP) with a frequency-flat dictionary~\cite{omp_exploit_sparsity, ip_omp}.
	\item The OMP-based estimator with the frequency-dependent dictionary of Section~\ref{sec:proposed_ce}.
	\item The proposed GSOMP-based estimator and its \ac{CRLB}.
\end{itemize}
The NMSE metrics for the LS method and the CRLB are computed using~\eqref{eq:mse_ls_fulltraining} and~\eqref{eq:crlb} in the numerator of~\eqref{eq:nmse}, respectively. The NMSE attained by each scheme is depicted in~Fig.~\ref{fig:NMSEvsSNR}(\subref{fig:NMSEvsSNR_a}). As we observe, the NMSE of the LS method is prohibitively high since it scales linearly with the number of BS antennas. Likewise, the NBOMP exhibits a very poor performance since it neglects the spatial-wideband effect. Moreover, the OMP-based estimator fails to successfully recover the common support in the low SNR regime, hence resulting in significant estimation errors. On the other hand, the proposed GSOMP-based estimator accurately detects the common support of the channel gain vectors for all SNR values ranging from $-15$~dB to $10$~dB, and thus attains the CRLB.  

Next, we focus on the state-of-the-art of estimation techniques based on the \ac{OMP}. To this end, we distinguish the work in~\cite{ce_omp_mmwave}, which proposed a \textit{nonuniform} dictionary and an RF pilot beam design based on the \ac{DFT} for a narrowband system with \ac{ULAs}; henceforth, we will refer to this scheme as OMP-DFT. Here, we extend the said design to the UPA case with spatial-wideband effects, and compare it with our proposed method. As we see from~Fig.~\ref{fig:NMSEvsSNR}(\subref{fig:NMSEvsSNR_b}), the GSOMP outperfoms the OMP-DFT. The poor performance of the OMP-DFT stems from the fact that the dictionary and RF pilot beams become highly correlated for a large number of BS antennas and high SNR values. To see this, recall that the dictionary resembles a \ac{DFT} matrix. Consequently, the product of the DFT-based pilot combiner and the dictionary tends to have multiple close-to-zero columns, hence destroying the incoherence of the equivalent sensing matrix.

\subsubsection{Multi-Antenna User}
We now investigate how multiple user antennas affect the channel estimation performance at the \ac{BS}. In order to have a fair comparison between the single-antenna and multi-antenna user cases, we fix the total number of antennas to $N_B N_U = 160$, and consider an $20\times 20$-element UPA at the \ac{BS} and an $4$-element ULA at the user.\footnote{In this way, the overhead of partial training, $0.8 N_{B}N_U$, is kept fixed too.} For $\varphi\sim\mathcal{U}(-\pi/2,\pi/2)$, the continuous spatial frequency $\omega = 1/2\sin\varphi$ lies in the interval $[-1/2, 1/2]$. Thus, the user's dictionary consists of the spatial frequencies $\{\bar{\omega}(p) = p/G^u, p=-(G^u-1)/2,\dots,(G^u-1)/2\}$. The elements of the pilot \ac{RF} beamformers $\{\mathbf{v}_i\}$ are selected from the set $\{-1/\sqrt{N_U}, 1/\sqrt{N_U}\}$ with equal probability. 

\begin{figure}[t]
	\centering	
	\includegraphics[width=0.91\linewidth]{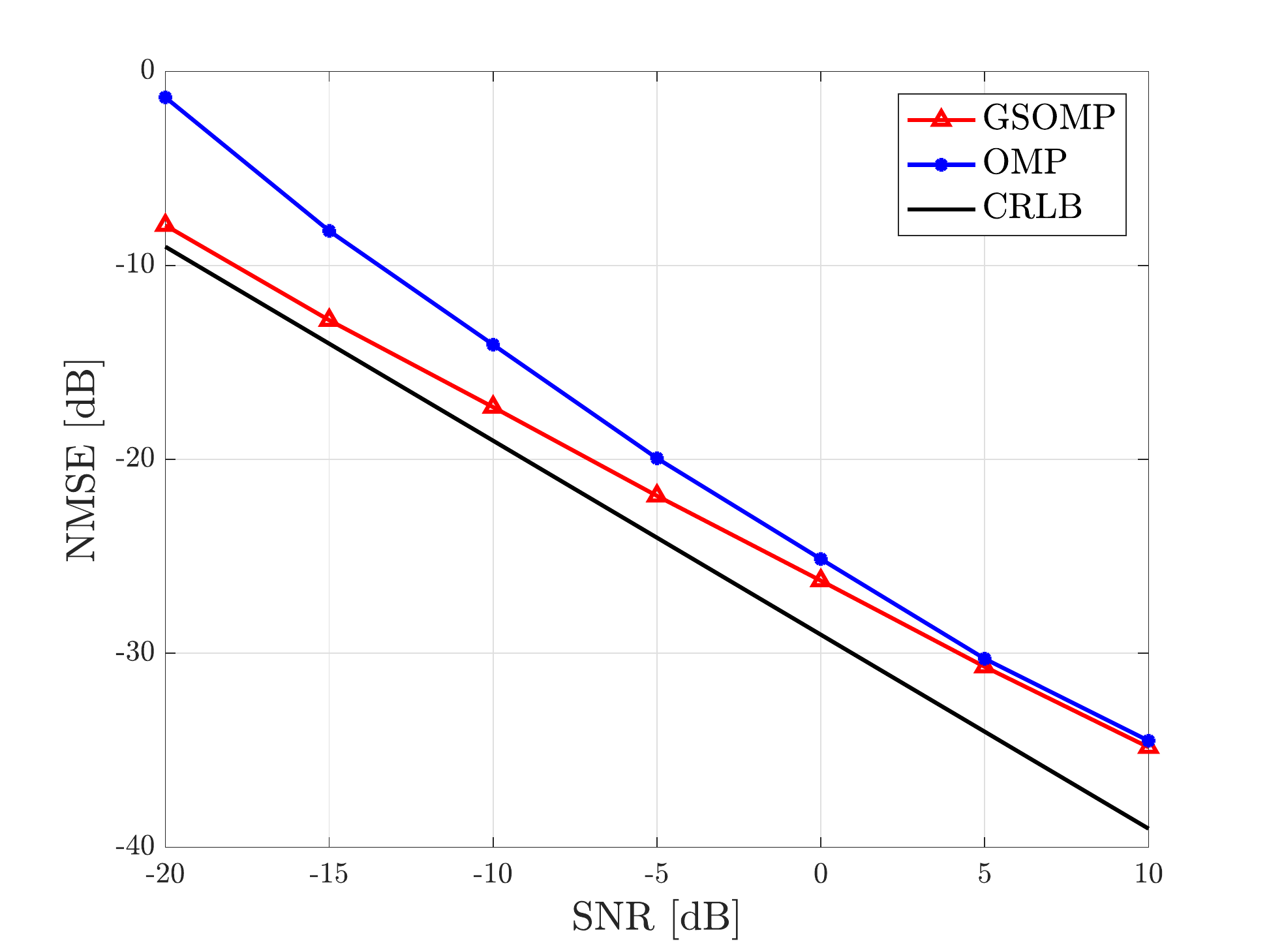}
	\caption{NMSE versus SNR for a user with an $4$-element ULA; $20\times 20$-element \ac{UPA}, $N_{\text{RF}}=2$, \ac{NLoS} channel with $L=3$ paths, $S = 400$ subcarriers, and super-resolution dictionaries with $G = 4N_B$ and $G^u = 4 N_U$.}
	\label{fig:NMSEvsSNR_MultiAntenna}
\end{figure}

The \ac{NMSE} is computed by replacing $\mathbf{h}[s]$ and $\hat{\mathbf{h}}[s]$ in~\eqref{eq:nmse} with $\text{vec}(\mathbf{H}[s])$ and $\text{vec}(\hat{\mathbf{H}}[s])$, respectively. The MSE of the \ac{LS} scheme \eqref{eq:mse_ls_fulltraining} is the same as in the single-antenna user case since we have kept fixed the total number of antennas. Figure~\ref{fig:NMSEvsSNR_MultiAntenna} depicts the performance of the GSOMP and OMP. As observed, there is a slight increase in the \ac{NMSE} compared to the single-antenna user case, i.e., Fig.~\ref{fig:NMSEvsSNR}(\subref{fig:NMSEvsSNR_a}). Furthermore, this increase becomes significant in the high \ac{SNR} regime, but yet, the proposed estimator outperforms the \ac{OMP} for low and moderate \ac{SNR} values. The performance degradation is because  the equivalent sensing matrices $\{\mathbf{\Phi}_s\}^{S-1}_{s=0}$ have higher total coherence compared to the single-antenna user case, which is defined for each matrix $\mathbf{\Phi}_s$ as~\cite{ce_omp_mmwave}
\begin{equation}
\mu(\mathbf{\Phi}_s)\triangleq \sum_{i=1}^{GG^u}\sum_{j=1, j\neq i}^{GG^u} \frac{|\mathbf{\Phi}^H_s(i)\mathbf{\Phi}_s(j)|}{\|\mathbf{\Phi}_s(i)\| \|\mathbf{\Phi}_s(j)\|}.
\end{equation}
It is worth pointing out that different pilot beam designs might change the performance of the estimators, which hinges on the coherence of the equivalent sensing matrices $\{\mathbf{\Phi}_s\}^{S-1}_{s=0}$.

\subsubsection{Subcarrier Selection}
\begin{figure}[t]
	\centering	
	\includegraphics[width=0.91\linewidth]{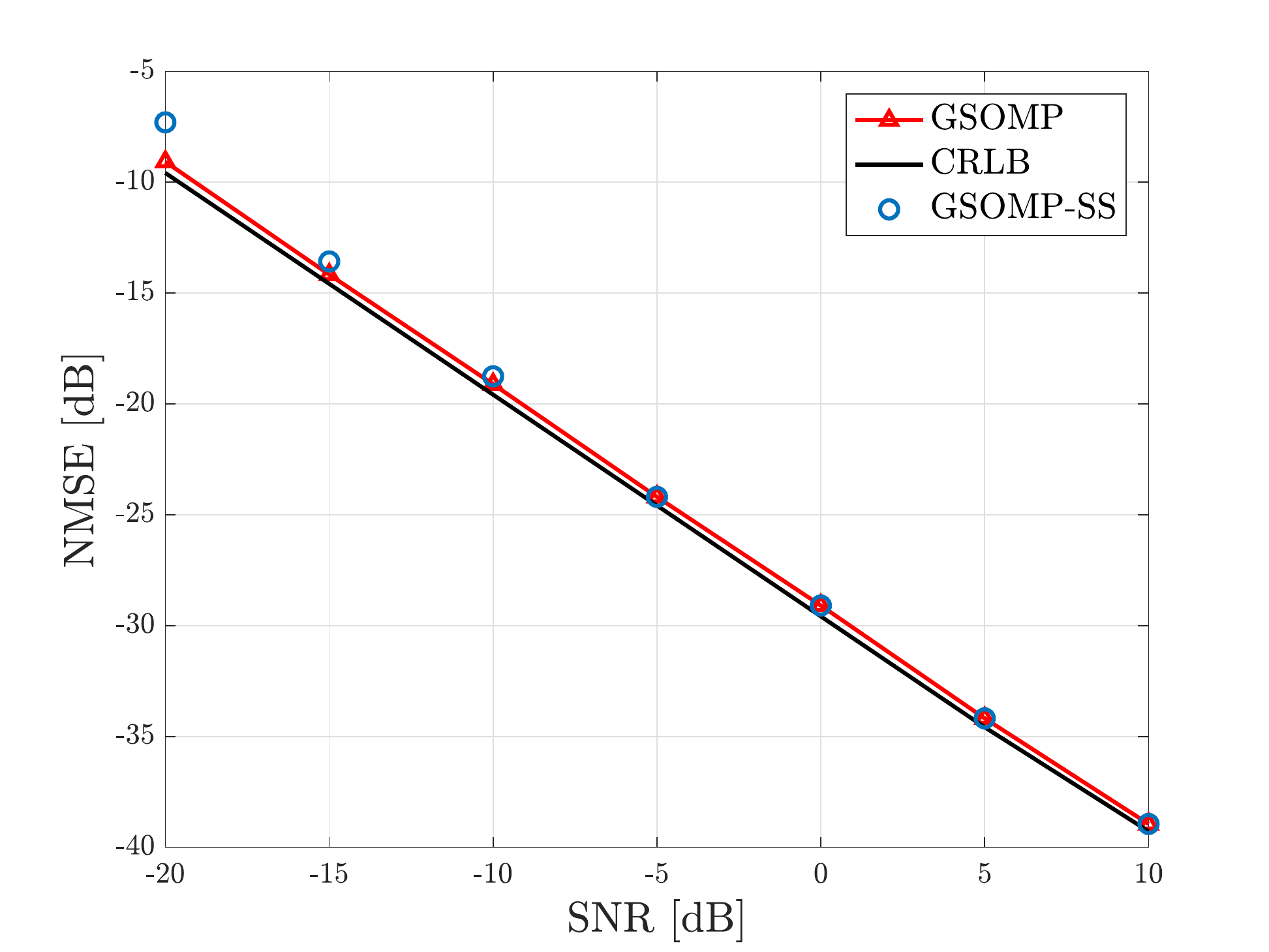}
	\caption{NMSE versus SNR for a single-antenna user. In GSOMP-SS, one pilot subcarrier per $50$ subcarriers is used to detect the common support; $40\times 40$-element \ac{UPA}, $N_{\text{RF}}=2$, \ac{NLoS} channel with $L=3$ paths, and $S=400$ subcarriers.}
	\label{fig:GSOMP_SS}
\end{figure}

In the previous experiments, we assumed that the \ac{GSOMP}-based estimator employs all the subcarriers, i.e., $|\mathcal{S}|=400$, to estimate the common support of the channel gain vectors $\{\mathbf{\beta}[s]\}_{s=0}^{S-1}$. However, this might lead to a very high computation burden. Thus, we can employ only a set of successive subcarriers to detect the common support, i.e., steps $2-8$ of Algorithm~$2$, and then use this support to estimate the channel at every subcarrier~$s\in\mathcal{S}$, which corresponds to step $9$ of Algorithm $2$. We refer to this scheme as GSOMP with subcarrier selection (GSOMP-SS). From Fig.~\ref{fig:GSOMP_SS}, we observe that we can accurately estimate the uplink channel in the moderate SNR regime by employing only a small number of pilot subcarriers in the common support detection steps. Note, though, that using one subcarrier per $50$ pilot subcarriers slightly increases the NMSE in the low SNR regime.  

\subsection{Hybrid Combining for Single-Antenna Users}

\subsubsection{Achievable Rate with Perfect CSI}
We start the performance assessment of our combiner by considering a \ac{LoS} channel. In this case, the complex path gain is given by $\beta_0(f) = \alpha_0(f)e^{-j2\pi f_c\tau_0}$, where $\tau_0 =~\mathtt{D}/c$ is the \ac{ToA} of the \ac{LoS} path, and $\alpha_0(f)$ is specified according to~\eqref{eq:path_atten_LoS}. For each channel realization, perfect knowledge of the DoA is assumed at the BS, which can be acquired using the GSOMP estimator. We also consider the following cases:
\begin{itemize}
	\item A fully-digital architecture where the BS employs the frequency-selective combiner $1/\sqrt{N_B}\mathbf{a}(\phi_0,\theta_0, f)$.
	\item A hybrid architecture where the \ac{BS} uses the narrowband combiner $1/\sqrt{N_B}\mathbf{a}(\phi_0,\theta_0, 0)$.
	\item A hybrid architecture where the proposed combiner~\eqref{eq:ttd_bf_upa} is used, with $N_{\text{sb}} = 10$ and $M_{\text{sb}} =10$ virtual subarrays. 
\end{itemize}
\begin{figure}[t]
	\centering	
	\includegraphics[width=0.92\linewidth]{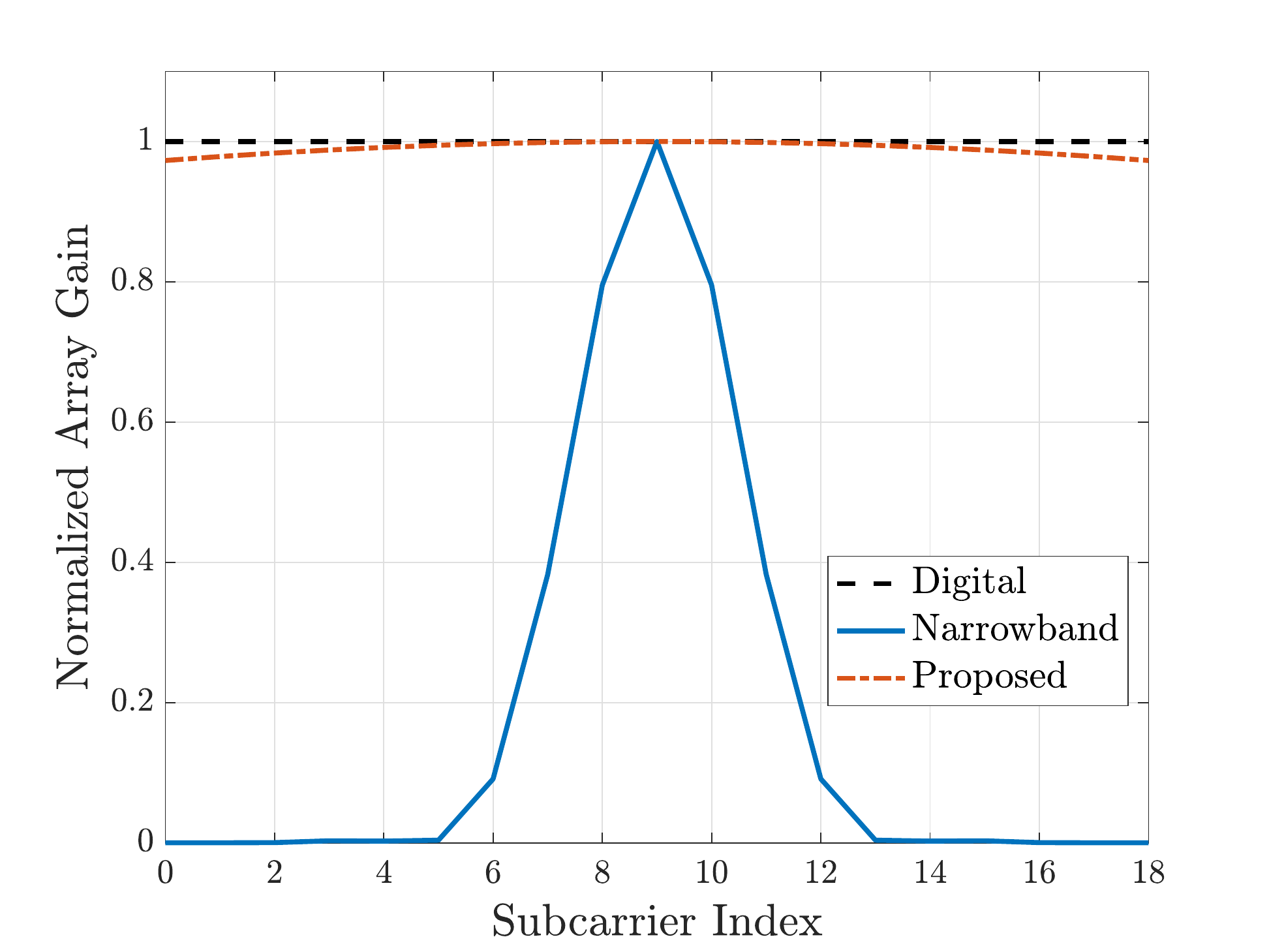}
	\caption{Normalized array gain for an $100\times 100$-element UPA. In the proposed scheme, $N_{\text{sb}}M_{\text{sb}} - 1 = 99$ \ac{TTD} elements are employed; \ac{LoS} channel, $(\phi_0,\theta_0) = (\pi/4, \pi/3)$, and $S=18$ subcarriers.}
	\label{fig:normalized_ag}
\end{figure}
The normalized array gain is plotted in Fig.~\ref{fig:normalized_ag}, where we see that our combiner atttains approximately the maximum gain over the entire signal bandwidth of $B = 40$ GHz. Next, we focus on the average achievable rate, which is calculated as  
\begin{equation}
	R = \sum_{s=1}^S \Delta B\mathbb{E}\left\{\log_2\left(1 + \frac{P_d |\mathbf{f}_{\text{RF}}^H\mathbf{h}[s]|^2}{\Delta B\sigma^2}\right)\right\},
\end{equation}
where $P_d = P_t/S$ is the power per subcarrier, and $\mathbf{f}_{\text{RF}}$ denotes the corresponding combiner.
\begin{figure}[t]
	\centering	
	\includegraphics[width=0.88\linewidth]{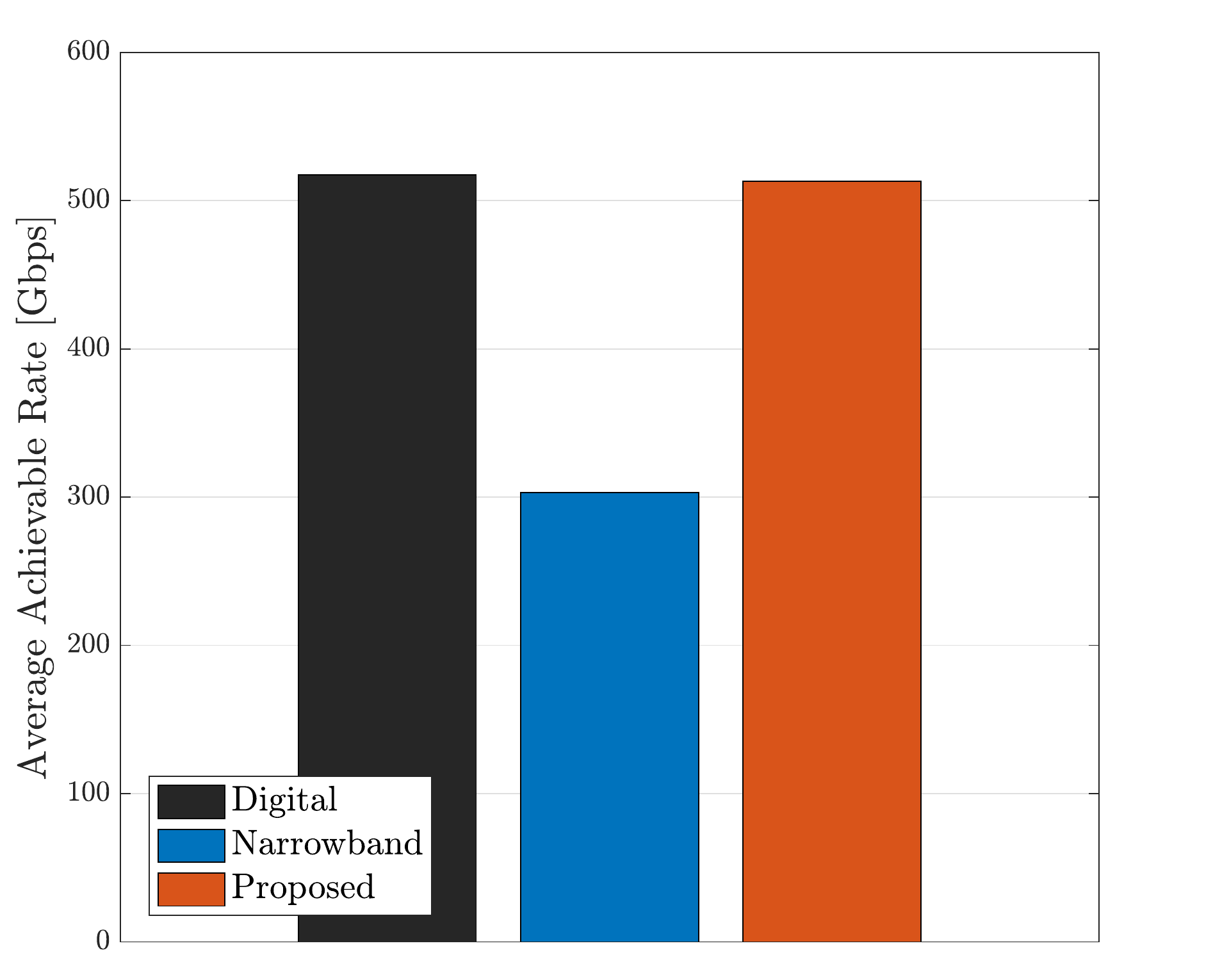}
	\caption{Average achievable rate under perfect CSI for a \ac{LoS} channel; single-antenna user, $100\times 100$-element UPA, $99$ \ac{TTD} elements in the proposed scheme, and $S=18$ subcarriers.}
	\label{fig:data_rate_los}
\end{figure}
The results are given in Fig.~\ref{fig:data_rate_los}. Specifically, the achievable rates are $517$~Gbps, $514$~Gbps, and $303$~Gbps for the digital, proposed, and narrowband schemes, respectively. Thus, the proposed combiner performs very close to the fully-digital scheme, while offering a $40\%$ gain with respect to the narrowband combiner. Additionally, this is done by employing only $N_{\text{sb}}M_{\text{sb}} - 1 = 99$ \ac{TTD} elements for an $100\times 100$-element \ac{UPA}, which yields an excellent trade-off between hardware complexity and performance. Lastly, note that transmission rates at least $R=0.5$ Tbps at $\mathtt{D}=15$ meters can be achieved through an $100\times 100$-element UPA, which would not be feasible with an equivalent ULA under a footprint constraint.

\subsubsection{Achievable Rate with Imperfect CSI} 
We now evaluate the average achievable rate attained by the proposed combiner along with the \ac{GSOMP}-based estimator. To this end, we consider a \ac{NLoS} multi-path channel. The complex path gain of the $l$th \ac{NLoS} path is $\beta_l(f) = \alpha_l(f)e^{-j2\pi f_c\tau_l}$, where $\tau_l$ is the \ac{ToA}, and $\alpha_l(f)$ is calculated according to~\eqref{eq:path_atten_NLoS} assuming $\phi_{i,l}\sim\mathcal{U}(-\pi/2,\pi/2)$. Under imperfect CSI, the BS treats the channel estimate as the true channel, and combines the received signal with the maximum-ratio combiner $\hat{\mathbf{h}}[s]/\|\hat{\mathbf{h}}[s]\|$. Let $\mathbf{h}[s] = \hat{\mathbf{h}}[s] - \mathbf{e}[s]$, with $\mathbf{e}[s]$ denoting the channel estimation error for the $s$th subcarrier. The combined signal for the $s$th subcarrier is then written as
\begin{align}\label{rx_signal_icsi}
y[s] &= \sqrt{P_d} \|\hat{\mathbf{h}}[s]\|x[s] - \sqrt{P_d}\frac{\hat{\mathbf{h}}^H[s]\mathbf{e}[s]}{\|\hat{\mathbf{h}}[s]\|}x[s] + \frac{\hat{\mathbf{h}}^H[s]}{\|\hat{\mathbf{h}}[s]\|}\mathbf{n}[s] \nonumber\\
& = \sqrt{P_d} \|\hat{\mathbf{h}}[s]\|x[s] + n_{\text{eff}}[s],
\end{align}
where $n_{\text{eff}}[s] =  (- \sqrt{P_d}\hat{\mathbf{h}}^H[s]\mathbf{e}[s]x[s] +\hat{\mathbf{h}}^H[s]\mathbf{n}[s])/\|\hat{\mathbf{h}}[s]\|$ is the effective noise. Unfortunately, it is challenging to derive an achievable rate of channel model~\eqref{rx_signal_icsi}  since the effective noise is correlated with the desired signal. Nevertheless, as shown in the previous numerical results, the channel estimation error is small. Hence, it is reasonably assumed that, conditioned on the channel estimates, the effective noise is uncorrelated with the desired signal. Then, we obtain the following approximation for the equivalent \ac{SNR} at the $s$th subcarrier~\cite{mMIMO_book2}
\begin{equation}
\text{SNR}_{\text{eq}}[s] \approx \frac{P_d \|\hat{\mathbf{h}}[s]\|^2}{\Delta B\sigma^2 + P_d\hat{\mathbf{h}}^H[s]\mathbf{R}_{\mathbf{e}[s]}\hat{\mathbf{h}}[s]/\|\hat{\mathbf{h}}[s]\|^2},
\end{equation}
where $\mathbf{R}_{\mathbf{e}[s]} \triangleq \mathbb{E}\{\mathbf{e}[s]\mathbf{e}^H[s]\}$. The corresponding average achievable rate under imperfect CSI is then~\cite{mMIMO_book2}
\begin{equation}
R \approx \sum_{s=1}^S \Delta B \mathbb{E}\left\{\log_2\left(1 +\text{SINR}_{\text{eq}}[s]\right)\right\}.
\end{equation}
A closed-form expression for $\mathbf{R}_{\mathbf{e}[s]}$ can be derived by assuming perfect recovery of the common support of the channel gain vectors. More specifically, from the CRLB analysis, we have that the error $\mathbf{e}[s] \triangleq\bar{\mathbf{A}}_s(\mathcal{I})\left(\hat{\bar{\bm{\beta}}}[s] - \tilde{\bar{\bm{\beta}}}[s]\right)$ is distributed as $\mathcal{CN}\left(\mathbf{0}, \mathbf{R}_{\mathbf{e}[s]} \right)$, where $\mathbf{R}_{\mathbf{e}[s]} = \bar{\mathbf{A}}_s(\mathcal{I})\mathbf{I}^{-1}\left(\tilde{\bar{\bm{\beta}}}[s]\right)\bar{\mathbf{A}}^H_s(\mathcal{I})$.
\begin{figure}[t]
	\centering	
	\includegraphics[width=0.88\linewidth]{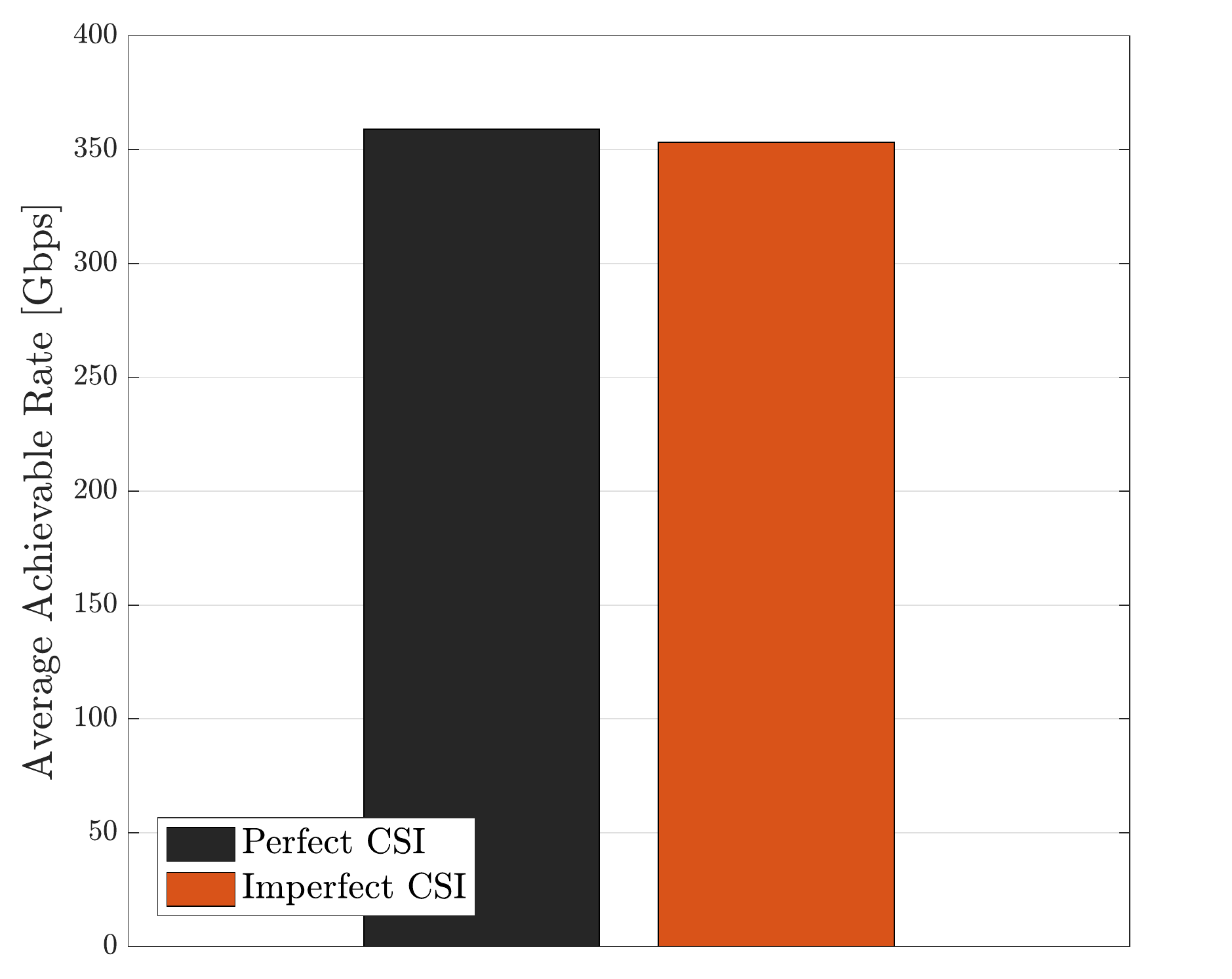}
	\caption{Average achievable rate under imperfect \ac{CSI} for a \ac{NLoS} channel with $L=2$ paths; single-antenna user, $100\times 100$-element \ac{UPA}, $99$~\ac{TTD} elements per RF chain, and $S=400$ subcarriers.}
	\label{fig:achievable_rate_icsi}
\end{figure}
Figure~\ref{fig:achievable_rate_icsi} depicts the average achievable rate under perfect and imperfect \ac{CSI}. In the imperfect CSI case, the common support of the channel gain vectors is computed by the \ac{GSOMP}-based estimator. As observed from Fig.~\ref{fig:achievable_rate_icsi}, the average achievable rate attained by the proposed channel estimator approaches that of the perfect \ac{CSI} case.

\subsection{Hybrid \ac{SVD} Transmission for Multi-Antenna Users}
\begin{figure}[t]
	\centering	
	\includegraphics[width=0.88\linewidth]{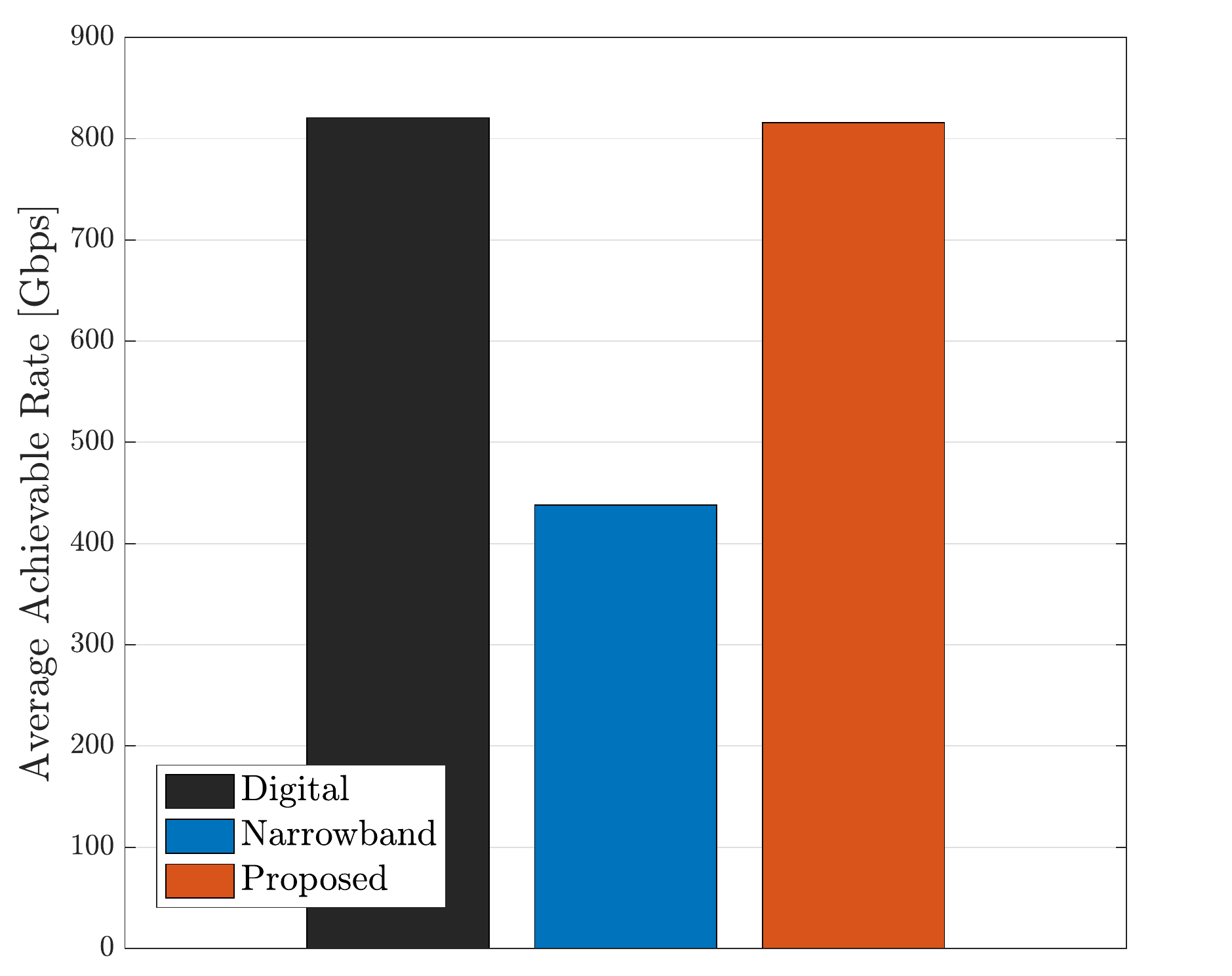}
	\caption{Average achievable rate a for a \ac{NLoS} channel with $L=2$ paths; multi-antenna user with an $2$-element ULA, $100\times 50$-element \ac{UPA}, $49$~\ac{TTD} elements per RF chain, and $S=400$ subcarriers.}
	\label{fig:svd_tx}
\end{figure}

In this section, we consider a multi-antenna user. As previously shown, we can accurately estimate the channel using the GSOMP-based estimator, and hence perfect \ac{CSI} is assumed. To have a fair comparison between the single-antenna and multi-antenna user cases, we fix the number of antennas to $N_UN_B = 100\times 100$, and we consider an $100\times 50$-element \ac{UPA} at the \ac{BS} and an $2$-element ULA at the user. Due to the small user array size, we assume a fully-digital array at the user, where $N^u_{\text{RF}} = N_U=2$. Subsequently, we compare the following transmission schemes:
	\begin{itemize}
		\item Digital: the combiner $\mathbf{F}[s]$ and precoder $\mathbf{B}[s]$ are designed using the \ac{SVD} of the channel $\mathbf{H}[s]$.
		\item  Proposed: the wideband \ac{RF} combiner $\mathbf{F}_{
			\text{RF}}[s]$ implements the scaled matrix $1/\sqrt{N_B}\mathbf{H}_{B}(f)$, defined in~\eqref{eq:H_b}, using \ac{TTD} and virtual array partition. The baseband combiner $\mathbf{F}_{\text{BB}}[s]$ and precoder $\mathbf{B}[s]$ are then designed using the \ac{SVD} of the effective channel $\mathbf{F}^H_{\text{RF}}[s]\mathbf{H}[s]$.
		\item Narrowband: the frequency-flat \ac{RF} combiner $\mathbf{F}_{
			\text{RF}}$ implements the scaled matrix $1/\sqrt{N_B}\mathbf{H}_{B}(0)$ defined in~\eqref{eq:H_b}. The baseband combiner $\mathbf{F}_{\text{BB}}[s]$ and precoder $\mathbf{B}[s]$ are then designed based on the \ac{SVD} of the effective channel~$\mathbf{F}^H_{\text{RF}}\mathbf{H}[s]$.
	\end{itemize}
	The average achievable rate is calculated as
	\begin{align}\label{eq:achievable_rate_wideband}
	R = \sum_{s=0}^{S-1}\sum_{n=0}^{N^u_{\text{RF}}}\Delta B\mathbb{E}\left\{\log_2\left(1 + \frac{p_{n,s}\sigma^2_n(\mathbf{F}^H[s]\mathbf{H}[s]\mathbf{B}[s])}{\Delta B\sigma^2}\right)\right\},  
	\end{align}
	where the set $\{p_{n,s}\}$ of powers is calculated using the waterfilling power allocation algorithm, and $\sigma_n(\cdot)$ denotes the $n$th singular value of the input matrix. From Fig.~\ref{fig:svd_tx}, we consolidate that effectiveness of the proposed \ac{TTD}-based method, which performs close to the fully-digital transmission scheme. More importantly, the deployment of a few antennas at the user side along with waterfilling power allocation boosts the average achievable rate compared to the single-antenna user case, which enables rates much higher than $R=0.5$ Tbps at a distance $\mathtt{D} = 15$~m. Another benefit of having multiple user antennas is the reduction of the \ac{BS} array size, which permits combating the spatial-wideband effect with a small number of \ac{TTD} elements. In particular, for the $100\times 50$-element \ac{UPA} under consideration, we have used $N_{\text{sb}} = 10$ and $M_{\text{sb}} = 5$ virtual subarrays, resulting in $N_{\text{sb}}M_{\text{sb}}-1= 49$ \ac{TTD} elements. 
	\subsection{Near-Field Considerations}\label{sec:spherical_wave}
\begin{figure}[t]
	\centering	
	\includegraphics[width=0.92\linewidth]{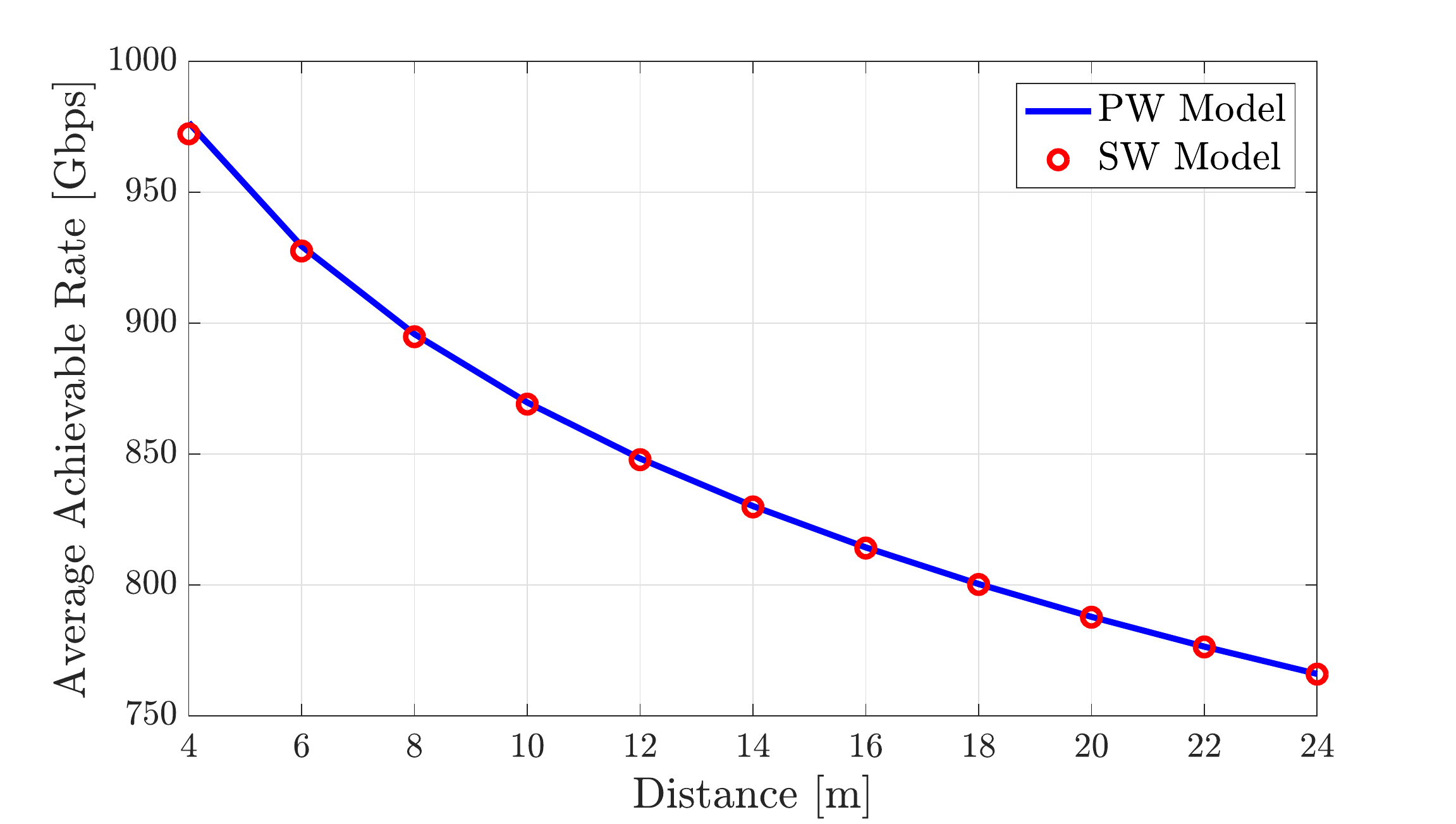}
	\caption{Average achievable rate of the \ac{TTD}-based wideband combiner for a \ac{LoS} channel; single-antenna user, and $100\times 100$-element \ac{UPA}.}
	\label{fig:pw_vs_sw}
\end{figure}

	In the far-field region, the spherical wavefront degenerates to a plane wavefront, which allows the use of the parallel-ray approximation to derive the array response vector~\eqref{eq:upa_response}. Due to the large array aperture of \ac{THz} massive MIMO, though, near-field considerations are of particular interest. Recall that near-field refers to distances smaller than the Fraunhofer distance $\mathtt{D}_f \triangleq 2 \mathtt{D}_{\max}^2/\lambda$, where $\mathtt{D}_{\max}$ is the maximum dimension of the antenna array, and $\lambda$ is the carrier wavelength. For a \ac{UPA} with $N=M$, we have $ \mathtt{D}_{\max}^2= 2 (N-1)^2 d^2$, i.e., length of its diagonal dimension, which leads to $\mathtt{D}_f = (N-1)^2\lambda$ for a half-wavelength spacing. Then, for $f_c = 300$ GHz and an $100\times100$-element \ac{UPA},  $\mathtt{D}_f \approx 9.8$ meters. As a result, the plane wave assumption may not hold anymore in small distances from the \ac{BS}~\cite{los_mimo}. In this case, a spherical wavefront is a more appropriate model~\cite{spherical_wave_model}. Under this model, the array response matrix, $\mathbf{A}(\phi,\theta,f)\in\mathbb{C}^{M\times N}$, of the \ac{BS} is defined as
	\begin{equation}
	[\mathbf{A}(\phi,\theta,f)]_{m,n} \triangleq e^{-j2\pi(f_c + f)\frac{\mathtt{D}_{mn}(\phi,\theta)}{c}},
	\end{equation} 
	where $\mathtt{D}_{mn}(\phi,\theta) = \left((x-nd)^2 + (y-md)^2 + z^2\right)^{1/2}$ is the distance between the $(n,m)$th BS antenna and the scatterer with coordinates $(x,y,z)$; $x \triangleq \mathtt{D}\cos\phi \sin\theta$, $y \triangleq \mathtt{D} \sin\phi \sin\theta$, and $z\triangleq\mathtt{D}\cos\theta$, where $\mathtt{D}$ denotes the distance from the $(0,0)$th BS antenna. The array response vector is then obtained as $\mathbf{a}(\phi,\theta,f) = \text{vec}(\mathbf{A}(\phi,\theta,f))$. We now calculate the average achievable rate for the \ac{TTD}-based combiner~\eqref{eq:ttd_bf_upa}
	under the plane and spherical wave models. The combiner is designed assuming a plane wavefront in both cases. From Fig.~\ref{fig:pw_vs_sw}, a very good match between the two models is observed even for distances smaller than the Fraunhofer distance. Thus, the proposed combiner can be used at near-field distances without incurring a significant rate loss. However, we stress that a comprehensive study of the near-field effects under different array arrangements and sizes is left for future work.  
	
\section{Conclusions}\label{sec:conclusions}
We have proposed a solution to the channel estimation and hybrid combining problems in wideband \ac{THz} massive MIMO. Specifically, we first derived the \ac{THz} channel model with \ac{SFW} effects for a \ac{UPA} at the \ac{BS} and a single-antenna user. We then showed that standard narrowband combining leads to severe reduction of the array gain due to beam squint. To tackle this problem, we introduced a novel \ac{TTD}-based wideband combiner with a low-complexity implementation due to the virtual subarray rationale. We next proposed a \ac{CS} algorithm along with a wideband dictionary to acquire reliably the \ac{CSI} with reduced training overhead under the spatial-wideband effect. To study the performance of the proposed schemes, we derived the \ac{CRLB} and computed the achievable rate under imperfect {CSI}. We also extended our analysis to the multi-antenna user case, and conducted numerical results. 
	
Simulations demonstrated that our design provides nearly \textit{beam squint-free} operation, as well as enables accurate \ac{CSI} acquisition even in the low \ac{SNR} regime. Regarding the insights drawn from our study, the deployment of multiple antennas at the user can alleviate the spatial-wideband effect by reducing the \ac{BS}' array size, whilst keeping constant the total number of antennas. As a result, the \ac{TTD}-based wideband array can offer the power gain required to compensate for the very high propagation losses at \ac{THz} bands. Additionally, in the case of multi-path propagation, it has been shown that \ac{SVD}-based transmission can boost performance and permit rates more than half terabit per second over a distance of several meters. In conclusion, wideband massive MIMO is expected to be a key enabler for future \ac{THz} wireless networks.

Regarding future work, it would be interesting to study the performance of wideband \ac{THz} massive MIMO under hardware impairments, as well as investigate the beam tracking problem in high-mobility scenarios. Moreover, it would be interesting to compare OFDM with SC-FDE, and derive an analytical expression for the PAPR metric.

\section*{Appendix A}
For the normalized array gain, we have that
\begin{align*}
&\frac{|\mathbf{a}^H(\phi,\theta,0)\mathbf{a}(\phi,\theta,f)|}{N_B} = \\
&=\frac{|\left( \mathbf{a}_x(\phi,\theta,0)^H \otimes \mathbf{a}_y(\phi,\theta,0)^H\right) \left(\mathbf{a}_x(\phi,\theta,f) \otimes \mathbf{a}_y(\phi,\theta,f)\right)|}{NM}\\
& = \frac{|\left( \mathbf{a}_x(\phi,\theta,0)^H \mathbf{a}_x(\phi,\theta,f)\right)  \left(\mathbf{a}_y^H(\phi,\theta,0)\mathbf{a}_y(\phi,\theta,f)\right)|}{NM}.
\end{align*}
Then, it holds
\begin{align*}
\frac{|\mathbf{a}_x(\phi,\theta,0)^H \mathbf{a}_x(\phi,\theta,f)|}{N}  &= \frac{1}{N}\left| \sum_{n=0}^{N-1}e^{-j2\pi fn \frac{d}{c} \sin\theta\cos\phi}\right | \\
& =  \frac{1}{N}\left|\frac{1 - e^{-j2\pi f N \frac{d}{c} \sin\theta\cos\phi}}{1 - e^{-j2\pi f\frac{d}{c} \sin\theta\cos\phi}}\right |\\
& = \frac{1}{N}\left| \frac{\sin\left(N\pi f\Delta_x\right)}{\sin\left(\pi f\Delta_x\right)} \right|\\
& = D_N(2\pi f\Delta_x),
\end{align*}
where $\Delta_x =  \frac{d}{c} \sin\theta\cos\phi$. Likewise, we get
\begin{align*}
\frac{|\mathbf{a}_y(\phi,\theta,0)^H \mathbf{a}_y(\phi,\theta,f)|}{M} = D_M(2\pi f\Delta_y),
\end{align*}
where $\Delta_y =  \frac{d}{c} \sin\theta\sin\phi$, which yields the desired result.

\section*{Appendix B}
Using the identity $\mathbf{a}_x \otimes \mathbf{a}_y = \text{vec}\left(\mathbf{a}_y\mathbf{a}^T_x\right)$, we have 
\begin{align}
&\mathbf{A}(\phi,\theta,f) \triangleq \mathbf{a}_y(\phi,\theta,f)\mathbf{a}^T_x(\phi,\theta,f)\nonumber\\[0.2cm]
&=
\begin{bmatrix}
\mathbf{a}_{y,1}(\phi,\theta,f) \\
\vdots \\
\mathbf{a}_{y,M_{\text{sb}}}(\phi,\theta,f)
\end{bmatrix}
\left[\mathbf{a}^T_{x,1}(\phi,\theta,f),\cdots, \mathbf{a}^T_{x,N_{\text{sb}}}(\phi,\theta,f)\right] \nonumber\\[0.3cm]
& =  \begin{bmatrix}
\mathbf{A}_{11}(\phi,\theta,f) &  \cdots  & \mathbf{A}_{1N_{\text{sb}}}(\phi,\theta,f)\\
\mathbf{A}_{21}(\phi,\theta,f)  & \cdots  & \mathbf{A}_{1N_{\text{sb}}}(\phi,\theta,f)\\
\vdots  & \ddots  & \vdots  \\
\mathbf{A}_{M_{\text{sb}}1}(\phi,\theta,f)   &  \cdots  & \mathbf{A}_{M_{\text{sb}}N_{\text{sb}}}(\phi,\theta,f)
\end{bmatrix},
\end{align}
where $\mathbf{A}_{mn}(\phi,\theta,f)\triangleq \mathbf{a}_{y,m}(\phi,\theta,f)\mathbf{a}^T_{x,n}(\phi,\theta,f)$. We also have that
\begin{multline}\label{eq:optimal_bf_upa}
\mathbf{A}_{mn} (\phi,\theta,f) = \mathbf{a}_{y,m}(\phi,\theta,f)\mathbf{a}^T_{x,n}(\phi,\theta,f) \nonumber \\
= e^{-j2\pi (n-1)\tilde{N}(f_c + f)\Delta_x - j2\pi (m-1)\tilde{M}(f_c + f)\Delta_y}\mathbf{A}_{11}(\phi,\theta,f).
\end{multline}
Using the above relationships, we can write 
\begin{equation}
\mathbf{A}(\phi,\theta, 0) \odot \mathbf{T}[s] = \mathbf{v}_y \mathbf{v}_x^T,
\end{equation}
where 	
\begin{equation}
\mathbf{v}_x = \left[e^{-j2\pi (n-1)\tilde{N}(f_c + f)\Delta_x}\mathbf{a}_{x,1}(\phi, \theta,0)\right]_{n=1}^{N_{\text{sb}}},
\end{equation}
and 
\begin{equation}
\mathbf{v}_y = \left[e^{-j2\pi (m-1)\tilde{M}(f_c + f)\Delta_y}\mathbf{a}_{y,1}(\phi, \theta,0)\right]_{m=1}^{M_{\text{sb}}}.
\end{equation}
Now consider a path with array response $\mathbf{a}^H(\phi, \theta, f)$. Then,
\begin{align*}
&\mathbf{f}^H_{\text{RF}} \mathbf{a}(\phi, \theta, f)=\\
&= \frac{1}{\sqrt{N_B}}\text{vec}^H(\mathbf{A}(\phi,\theta,0) \odot \mathbf{T}[s])\mathbf{a}(\phi, \theta, f)  \\
 &= \frac{\sqrt{N_B}}{N_B}\left(\mathbf{v}^H_x\otimes \mathbf{v}^H_y\right) \left(\mathbf{a}_x(\phi,\theta,f) \otimes \mathbf{a}_y(\phi,\theta,f)\right)\\
 & = \frac{\sqrt{N_B}}{N_B}\left(\mathbf{v}^H_x\mathbf{a}_x(\phi,\theta,f)\right)\left(\mathbf{v}^H_y\mathbf{a}_y(\phi,\theta,f) \right)\\
 & = \sqrt{N_B} \frac{\mathbf{a}^H_{x,1}(\phi, \theta, 0)\mathbf{a}_{x,1}(\phi, \theta, f)}{\tilde{N}} \frac{\mathbf{a}^H_{y,1}(\phi, \theta, 0)\mathbf{a}_{y,1}(\phi, \theta, f)}{\tilde{M}}.
\end{align*}
As a result, we obtain~\eqref{eq:array_gain_proposition} in Proposition 1. 

\end{document}